\documentstyle[11pt,aasms4]{article}
\def\etal{{\it et al.\ }}

\def\putplot#1#2#3#4#5#6#7{\begin{centering} \leavevmode
\vbox to#2{\rule{0pt}{#2}}
\includegraphics{#1}

\end{centering}}

\begin{document}
\title{Far-Ultraviolet Imaging of the Field Star Population in the
Large Magellanic Cloud with HST{\footnote{Based on observations with the NASA/ESA
{\it Hubble Space Telescope}, obtained at the Space Telescope Science
Institute, which is operated by AURA, Inc., under NASA contract NAS 5-26555}}}


\author{Noah  Brosch{\footnote{Email: noah@wise.tau.ac.il}}\altaffilmark{, 3}, Michael Shara, John MacKenty, David Zurek and Brian McLean}
\affil{Space Telescope Science Institute \\ 3700 San Martin Drive \\
Baltimore MD 21218, U.S.A.}

\altaffiltext{2}{On sabbatical leave from the Wise Observatory and 
the School of Physics and Astronomy,
Raymond and Beverly Sackler Faculty of Exact Sciences,
Tel Aviv University, Tel Aviv 69978, Israel}


\begin{abstract}
We present an analysis of the deepest pure-UV observations 
with the highest angular resolution ever performed, a set of 
12 exposures with the HST WFPC2 and F160BW filter obtained in 
parallel observing mode, which cover $\sim$12 square arcminutes in the LMC,
North of the bar and in the ``general field'' regime of the LMC. 
The 341 independent measurements of 198 
objects represent an accumulated exposure of $\geq$2 10$^4$ sec 
and reveal stars as faint as m$_{UV}\simeq$22 mag. The observations
 show that $\sim$2/3 of the UV emission from
the LMC is emitted by our HST-detected UV stars in the field, {\it i.e.,
not} in clusters or associations.
We identified optical counterparts in the ROE/NRL
photometric catalog for $\sim$ 1/3 of the objects. 
The results are used to discuss the nature of these UV sources, to estimate
the diffuse UV emission from the LMC as a prototype of dwarf 
galaxies, and to evaluate the contamination by field
stars of UV observations of globular and open clusters in the LMC. We find
that the projected density of UV stars in the general field of the
LMC is a few times higher than in the Galactic disk close to the Sun.
Combining our data with observations by UIT allows us to define the
stellar UV luminosity function from m$_{UV}$=8 to 18 mag, and to confirm
that the field regions in the LMC have been forming stars at a steady
rate during the last 1 Gyr, with an IMF close to the Salpeter law.

\end{abstract}


\section{Introduction}

The Large Magellanic Cloud (LMC) is the nearest galaxy to the Milky
Way (with the exception of a merging dwarf galaxy), one in which individual
stars can be distinguished and studied fairly easily with instruments of the
highest angular resolution such as the Hubble Space Telescope (HST). The 
LMC is different from the Milky Way (MW) in that it shows more
intense star formation (SF), at least when compared with the Solar 
neighborhood, {\it i.e.,} the MW region within $\sim$2 kpc of the Sun where the
interstellar extinction is reasonably small and individual stars can be
easily studied in the visible region of the spectrum.

The difference in SF, coupled with the much smaller size of the LMC compared
with the MW, implies that this is a dwarf irregular or a tidally
truncated small spiral, in which the SF processes are probably different 
from those in the MW. Feitzinger \etal (1987) argued that the
LMC is a good example of a galaxy in which the SF proceeds
primarily via the stochastic self-propagating SF mechanism (SSPSF: 
Gerola \& Seiden 1978), whereas the conventional view ({\it e.g.,} 
Kaufman 1979) is that the SF in the MW is driven mainly by
spiral density waves. The different SF mechanism, together with the possibility
of studying individual stars, are the reasons the LMC is a very popular 
target for studies of SF-related phenomena ({\it e.g.,} Geha \etal 1998,
Battinelli \& Demers 1998).

There is good reason to study the population of field stars
in the LMC; these objects represent the unspectacular star formation
processes in the LMC and, by inference, in dwarf irregular galaxies in 
general. If we manage to understand the ``quiet'' mode of star formation, 
{\it i.e.,}
that which is not generated in a starburst, we may better understand
the origin of the UV radiation in the Universe. Note that in starburst 
galaxies only $\sim$20\% of the UV light at 2200\AA\, is produced
by stars in clusters (Meurer \etal 1995) and the majority of the light
originates from a general population of UV stars. This diffuse UV
emission might be due to the stars that have been ejected from regions 
of active SF. Studies of the field
star population of the LMC were reported by {\it e.g.,} Elson \etal (1997),
Geha \etal (1998). These were performed with HST in optical and 
near-infrared bands, being rather insensitive to massive/hot stars.
The stellar populations of the LMC were reviewed recently by Feast (1995),
who did not find strong evidence for a starburst triggered by a 
LMC-SMC-Milky Way interaction. He argued that the  SF
increase  in the LMC was caused by the collapse of the system to a 
plane about 4 Gyrs ago. 

The nature of the fainter ultraviolet (hereafter UV, covering the
approximate wavelength range $\lambda\lambda$1000-3000\AA\,) 
sources has not yet been resolved. This is in part due to the
fact that since the TD-1 
mission in the late 1960s (Thompson \etal 1978)
there has not been an all-sky survey in the UV. There has also 
not been an instrument capable of providing good imaging
in the UV, from which photometric parameters and shape information 
on extended objects could be derived. Previous deep UV observations
have been perfomed by the UIT Shuttle-borne telescope (Stecher \etal 1992)
and by the FOCA balloon-borne telescope (Milliard \etal 1992).
The latter are in a spectral band centered at 2015\AA\,
and $\sim188$\AA\, wide; the width of the band depends essentially on
the altitude of the balloon which carried the FOCA telescope to 42+ km
altitude. Because of this, the FOCA passband has a slight red leak
($\sim10^{-2}$),
which depends on the exact altitude of the instrument during the 
observation. The results from the diverse FOCA flights have not yet
been published in a systematic manner, as the ground-based follow-up
is still in its early phases. The few
published FOCA results, supplemented by ground-based follow-up
observations from  large telescopes ({\it e.g.,} Treyer \etal 1998),
are sufficiently intriguing to warrant concerted attempts to observe
faint UV sources with good angular resolution. These results indicate
that $\sim$50\% of the UV sources in a field centered on the Coma
cluster are background galaxies ranging to z$\approx$0.7 with
strong emission lines. Similar results were announced by the
group headed by R. Ellis (Treyer \etal 1998).

Other UV imaging instruments, such as S201 (Carruthers \etal 1977) or
FAUST (Bowyer \etal 1995), had much lower angular resolution than
either UIT ($\sim$3 arcsec) or FOCA ($\sim$10 arcsec) and did not
reach objects as faint as these. However, these low angular resolution
imagers offer unique panoramic views of the LMC in the UV. 
Previous UV imaging observations of the LMC were 
reported by Page \& Carruthers  (1981) using the S201 camera carried to 
the Moon's surface during the Apollo 16 flight; by Court\'{e}s \etal 
(1984) with the wide-field camera operated on the Space Shuttle; by Smith 
\etal (1987) using a rocket-borne UIT prototype; and by Court\'{e}s 
\etal (1995) with the FAUST instrument. In addition, observations in the 
UV near 30 Dor with photometers were reported by Koornneef (1977) using the 
ANS satellite with five UV bands and with an angular resolution of 2.5 arcmin; 
by Nandy \etal (1979) and Morgan \etal (1979) from the TD-1 satellite with 
four bands and with 1.7 arcmin resolution; and on stellar assocations in the 
LMC with the Skylab S183 experiment (Vuillemin 1988). Most LMC imaging in 
the UV was done with very wide field cameras and with low angular resolution, 
of order 3-5 arcmin. Despite the low resolution, which allows only for the 
mapping of the general UV emission pattern, these observations generated 
quantitative measures of the emission from unresolved sources and diffuse
nebulosity in the LMC. In particular, Page \& Carruthers (1981) 
measured the total UV emission from the LMC to be S$_{1400}$=3.4 10$^{-7}$
erg sec$^{-1}$ cm$^{-2}$ \AA\,$^{-1}$ ster$^{-1}$ (at 1400\AA\,) and 
S$_{1300}$=3.8 10$^{-7}$ erg sec$^{-1}$ cm$^{-2}$ \AA\,$^{-1}$ ster$^{-1}$
(at 1300\AA\,).

The 30 Doradus and SN 1987A regions in the LMC were observed with higher 
angular resolution by the UIT and the results were
published by Cheng \etal (1992). The latter observation is relevant, because
it relates to a region of the LMC very close to that studied here. Similar
UIT observations, of the associations LH 52 and LH 53, and the SNR N49 in the
LMC, were reported by Hill \etal (1995). The field population of
UV stars was studied by Hill \etal (1994), who found a lower star
formation (SF) rate there than in the stellar associations of the LMC, 
as well as a SF less biased toward high-mass objects. Finally, all the
UV sources in the direction of the LMC which have been observed by UIT have
been collected in a catalog by Parker \etal (1998).
The results from the two UIT flights constitute the 
deepest catalog of UV objects in the LMC published to date (Parker
\etal 1998). The UIT observations yielded photometry from images
with reasonable resolution: FWHM=3.4 arcsec. A comparison of the
UIT-derived magnitudes and synthetic photometry for 255 stars in
common with the IUE observatory show no systematic shift, but a
standard deviation of 0.25 mag in the distribution of magnitudes. 

It is possible that many UV sources are cosmologically-important objects,
as Treyer \etal (1998) have shown. These sources may account for a large part 
of the
extragalactic diffuse UV background (dUVB); this may not vanish completely in
directions with very low HI column densities, indicating that not all the
dUVB is dust-scattered Galactic UV light (Leinert \etal 1998). This motivated
us to study the deepest UV observations obtained in parallel mode
by the largest orbiting
astronomical telescope, the Hubble Space Telescope.
HST is equipped, in principle, with three UV-imaging instruments:
WFPC2, FOC, and STIS. The second instrument has an extremely restricted
field of view, while the third has a similarly small field but a much
higher photometric accuracy, dynamic range, and throughput than the
FOC. However, STIS cannot be used (yet) for parallel UV imaging 
observations because of the 
possible damage to the MAMA detectors by UV sources, and most of 
its pointings are in the direction of crowded,
well-defined fields such as galactic nuclei and globular clusters
where the sources are fainter than the damage limit.
It is clear that, at present, only WFPC2 can offer a reasonably
large field combined with some UV throughput for the serendipitous study 
of the UV sky.

A number of papers have been published which analyse data collected with
the WFPC2 UV filter set. Most of these filters have significant red leaks
and the accuracy of the photometry cannot always be fully estimated. For this 
reason, many researchers prefer to use the F160BW filter. This ``Wood's filter'' 
separates a pure UV band with an effective wavelength $\lambda_{e}$=1491\AA\,
and bandpass 446\AA\, (Holtzman \etal 1995: H95b). The  
F160BW filter has a very low throughput (approximately one order of
magnitude below other UV filters from the WFPC2 complement); however, its bandpass 
has no red leak, unlike other WFPC2 filters. It also vignettes 
$\sim$1/8  of the field imaged through each WF CCD (the PC chip is not
vignetted). More details about this filter can be found
in Watson \etal (1994).
Observations using F160BW to study hot stars in globular 
clusters were reported by {\it e.g.,} Mould \etal (1996) and Cole \etal 
(1997). Hunter \etal (1997) studied the UV emission from another cluster 
(R136) with the F170W filter, as did Gilmozzi \etal (1994) for the ``double 
cluster'' NGC 1850A+B; the F170W filter has a very strong red leak and the 
interpretation of the results with it is difficult. 
 
In order to realize fully the UV imaging potential of the WFPC2+F160BW
combination, a program of parallel imaging was initiated by JM. Some
archived observations obtained in this program are used here to 
study the field population of
UV stars in the LMC. We use the results to estimate the diffuse UV emission
from the LMC and by inference, from other star-bursting dwarf galaxies,
and to evaluate the degree of contamination by field UV stars of the
pointed observations of globular and open clusters in the LMC.
Our observations and reductions are described in Section 2 and the 
photometry is discussed in Section 3. Optical counterparts are 
identified in Section 4, and their nature is discussed in Section 5.
The detection of a significant number of UV stars allows us to draw some 
general conclusions about the SF in the LMC.

\section{Observations}

The HST observed objects in the N132D region of the LMC with the
Faint Object Spectrograph from 23 to 26 August 1995 collecting data 
for proposal 5607. During these
very long pointings the WFPC2 was commanded to observe in parallel
with the F160BW filter.  These observations are ideal to study
the population of hot stars in the crowded environment of the
LMC. During the LMC session the WFPC2 acquired 12 images with F160BW 
with exposure times of 2300 sec to 4900 sec, which 
we analyze here. A log of the HST data sets used in this paper is 
given in Table 1.

The region observed by HST is centered at (J2000) 5$^h$ 25$^m$, 
--69$^{\circ}$ 30'; the foreground extinction in this direction
is E(B--V)$\leq$0.1 (Lucke 1974). The area is located near the 
southeast end of the LMC bar, not on the bar itself but rather on its
northern edge, $\sim0^{\circ}.7$ off its major axis, and
$\sim1^{\circ}.3$ off the LMC's kinematic center. The list of UV-bright 
objects or
regions in the LMC detected by S201 (Page \& Carruthers 1981) contains no
object in this direction, and neither does the catalog of infrared sources in 
the LMC and SMC (Schwering \& Israel 1990). The map of radio continuum 
produced by
Xu \etal (1992) shows that at the location of the HST WFPC2 pointing
there are no strong ridges of 6.3 cm continuum emission. The map of diffuse
H$\alpha$ emission (Fig. 8 in Xu \etal 1992) shows that the line emission
is also minimal in this location, but that 30 Dor, about half a degree to 
the East,
is very bright. The region is one of those studied by Davies \etal (1976)
with UK Schmidt plates through a 100\AA\, wide H$\alpha$ filter to search
for  nebulosities. The N132D region is noted in their Table II along with
N132H as being faint and having ``knots in envelope'', a size of 7$\times$4 
arcmin, and corresponding  to the MC39 11cm radio source. MC39 is a radio
and Xray-detected supernova remnant (Clark \etal 1982), located $\sim$10 arcmin 
south of the WFPC2 region imaged in
the UV. The wide field imaging by the S183 experiment indicates that the 
emission at 2600\AA\, (250\AA\, passband) from the location of the HST pointings analyzed 
here is less than 2.69 10$^{-14}$ erg sec$^{-1}$ cm$^{-2}$ \AA\,$^{-1}$ 
(square arcmin)$^{-1}$ (Vuillemin 1988, Fig. 1b). This figure also shows that the 
HST pointing was not towards any of the UV-bright stellar associations in the LMC.

The image of the N132 field in Fig. 3 of Parker \etal (1998) contains the HST
field studied here in its lower-left (north of the center) part. No exceptionally 
high stellar density is apparent, nor is this a very sparse field. A careful
comparison between the 11$^{\circ}$.4 diameter image obtained
by Smith \etal (1987) at 1590\AA\, and shown as the top
figure of their Plate 23 and the location of the N132 field in the R-band
image shown by Parker \etal (1998, Fig. 2) emphasizes that the location
of the HST region studied here is off the LMC bar. Although the diffuse
UV emission in the vicinity of N132 seems to be higher than from 
the rest of the LMC disk, it is clear that the HST observations pointed to an 
area of unexceptional (faint) diffuse UV emission, between Shapley Constellations 
II and V (see Fig. 4 of Smith \etal 1987). As it is not clear how good is the
flat fielding of the very wide field image in Smith \etal (1987), we cannot
remark on the absolute UV surface brightness from the location of HST
observations.

The region observed with the HST is therefore typical of the ``field'' 
stellar population of the LMC,
{\it i.e.,} not containing stellar associations or clusters, HII regions or 
known supernova remnants. It is, thus, representative of the field stellar 
population in the nearest dwarf galaxy which shows intense star formation.  
The LMC images obtained by HST are shown in the mosaic of Figure 1, where the 
different frames have been combined using the astrometric parameters of 
the PC chip from set A as produced by the HST reduction pipeline. The other
images were linked into the mosaic through stars in common with previously
linked images. We indicate the scale of the figure with a 30 arcsec bar and 
the north and west directions by vectors plotted at the figure's lower right 
corner. The mosaic is shown here for display purposes only and it was not used in
the photometric reduction. Note also that in its creation we made no attempt 
to remove distortions, which are small, in general, for the WFPC2.

The images are grouped by coordinates in sets (A through F), which 
were obtained in the same sky position, with the same guide 
stars, and at the same HST roll angle. Using the task {\bf gcombine} of IRAF
(in stsdas.toolbox.imgtools), with the CCDCRREJ option which is based
on rejection of pixels higher by more than five standard deviations from
their immediate surroundings, or with the combination of images based on 
crrej (in stsdas.hst\_calib.wfpc) where pixels in stacked images are rejected 
by a similar algorithm, it is 
possible to eliminate most of the cosmic rays (CRs). These are especially
troublesome in very long WFPC2 exposures and it is virtually impossible to see
any UV source before combining at least two images and rejecting the CR events.

Even with the CCDCRREJ option enough traces of CR events remained in
the combined frame, mainly at locations where one CR track crossed
another, to significantly confuse the detection of faint objects. This is
because of the low throughput of the WFPC2+F160BW combination, the very low
UV sky background in this spectral region, and the susceptibility of the 
WFPC2 CCDs to CR events. However, these cases could be rejected by blinking 
the combined images against individual frames used in the same combination.
The case for dataset F is particularly difficult to analyze, as it consists 
of a single long exposure with significant but not identical overlap with 
set E. For this dataset we identified only objects visible in the original 
frames which had the proper intensity profile of a star (after some CR-cleaning
with the {\bf cosmicray} task in noao.imred.ccdred).

We used the combined frames to identify genuine UV sources by blinking combined
sets of frames with significant overlap. The objects recognized this way are 
listed with a double or triple entry in Table 2. We accepted only those
sources which appeared in a combined image, and which we could recognize on the two
(or three) original images which made up the combined image and 
had the ``proper'' shape parameters (FWHM$\simeq$0.2 arcsec) expected of genuine stellar images.
We calculated the celestial coordinates of all sources and performed
aperture photometry with the {\bf phot} package (in noao.digiphot.apphot) with a
0.5 arcsec round aperture. This is justified, given the PSF FWHM of images obtained
with F160BW (up to $\sim$0.26 arcsec; Watson \etal 1994). The {\bf phot} package
performs aperture photometry and subtracts the sky background as estimated
from a ring around the aperture. When performing aperture photometry, the
undersampling of the WF PSF is not important.

The photometry has not been corrected 
for charge transfer efficiency (Whitmore \& Heyer 1997), because we  found this 
to be insignificant in comparison with the other uncertainties in 
the data. For the same reason we did not correct for PSF variability.
We performed aperture corrections by subtracting 0.1 mag from the {\bf phot} results. 
We also corrected for the molecular contamination 
of the WFPC2 (MacKenty \etal 1995). The photometry is reported 
in Table 2 as {\it monochromatic magnitudes} at $\lambda$=1491\AA\,, with
\begin{equation}
m_{UV}=-2.5 \times log f_{1491} -21.1 
\end{equation}
and $f_{1491}=(\frac{DN}{sec}/GR_i) \times F_{\lambda_0}$, where
$GR_i$ is the gain ratio and $F_{\lambda_0}$ is the zero point
constant from Holtzman \etal (1995).  

We did not specifically exclude sources in the areas vignetted by the F160BW
filter. These are included in the Table and are marked by a {\bf V}
following the dataset and CCD chip identifiers given in column 6. Among the 
341 independent measurements, only 28 are of objects in vignetted locations
on the WF chips. In four cases a source with multiple measurements appears 
vignetted in one image but is not vignetted on others; this allows a comparison
of photometry which helps evaluate the effect of vignetting. With one exception
(source no. 45) the two photometry measurements agree to within 0.27 mag
or better. We conclude that the vignetting by the F160BW filter does not 
appear to significantly influence our photometry.

\section{Results}

We present  the raw UV photometry results for all the objects 
identified in the WFPC2 images in Table 2. Each independent measurement is
identified by a source number, which is the running number of UV sources 
and indicates the order in which the sources were identified.
Some objects appear in more than one  (combined) image; for each measurement 
we present the derived value and the photometric error, the image set from which 
the particular result originated (A, B, C, {\it etc.}), and the CCD chip in 
which the object appeared. For example, the first object in Table 2 was detected 
in the PC chip of the combined set A, at pixel coordinates (242.117; 260.523)
and its raw UV magnitude was 19.190$\pm$0.041.

The on-chip pixel coordinates are listed beside the celestial coordinates, 
to aid further
investigations of the sources. The correlation between different 
measurements was done through the celestial coordinates (calculated 
with the IRAF {\bf metric} task, in stsdas.hst\_calib.wfpc), recognizing 
two separate observations as
belonging to the same object if their image centers were separated by
$\leq$0.5 arcsec. The photometric error listed in Table 2 is 
that given by the {\bf phot} task in IRAF,
{\it i.e.,} a formal error depending on the photon count and the local
background; this does not reflect the calibration error of the
photometry from F160BW images or residual flat field errors. Note
that the F160BW flat field correction can be quite noisy (Biretta \& Baggett 1998).
The calibration error, on the other hand, is $\sim$10\% at the center of
the CCD chips (Baggett \etal 1998).

We combined the different measurements of the same object in a single
entry in Table 3, which has the source coordinates,  UV magnitudes
and errors. We identify a source by its first listed source number in 
Table 2. Figure 2 shows the distribution of remeasurements of the same 
object; approximately half of the objects have been measured more than once.
Figure 3 shows the distribution of apparent UV magnitudes from Table
3. Objects with more than one measurement are shown in the lower
panel. 

A formal comparison of the two magnitude distributions, {\it e.g.,} via
a Kolmogorov-Smirnov test of the cumulative distributions,
shows that they are not drawn from the same parent population. This is 
probably the combination of differences introduced at the faint
end of the distribution by a few spurious sources and at the bright end
by the differnt areal coverage; while singly-detected sources can
appear anywhere on the mosaic in Fig. 1, the multiply-measured objects
{\bf must} reside in the region of image overlap, which is
smaller than the total area observed by HST. Note also that
faint sources may register on an individual image but not on
another from a different image combination; it  would contribute
a real source measured only once, although it may reside in an
overlap region.

Fig. 3 shows that (a) brighter objects are more frequently seen in
multiple exposures, and (b) despite the caveat in the previous paragraph,
the faint end of the distribution is not dominated by the singly-measured 
objects. In fact, the drop in the  number-magnitude plots
occurs in both histograms at m$_{UV}$=19, indicating a similar degree of
completeness for the singly-measured and multiply-measured objects. It is 
possible that a few faint sources are spurious, but most represent real objects
conservatively chosen while adopting stellar images as candidate objects. We  
have likely missed some fainter LMC objects through this methodology,
but we are fairly certain that we excluded virtually all spurious sources.
The number-magnitude diagrams indicate that our photometry becomes
incomplete at m$_{UV}\simeq$18.5.

Comparing the independent measurements of the same objects, after correcting 
for WFPC2 contamination, we derive an unbiased estimator of the measurement 
error, which disregards calibration unknowns but which is more representative 
of the internal photometry error than the value given by {\bf phot}. This is 
the measurement error we adopt and present in Table 3. We show in Figure 4
the distribution of the individual FUV measurements for objects
which have been measured independently twice (filled squares) where the
error bars are the formal {\bf phot} errors of these points, and the 
average value adopted in Table 3 (empty diamonds) with its error bar estimated 
from the dispersion of the individual measurements. It is clear that the 
dispersion of the measurements is reasonably small for objects with
 m$_{UV}\leq$19, but increases for objects at  m$_{UV}\geq$19 to one
or more magnitudes. This demonstrates once again that the completeness limit 
of the UV observations reported here is near m$_{UV}\approx$18.5.

We note that none of the $\sim$200 UV objects identified here appear
convincingly extended when examined on the HST F160BW images. We 
conclude tentatively that most of them are probably stars or starlike
objects, and that in the direction of the small LMC area sampled by
the observations there are no UV-bright background galaxies.


The Ultraviolet Imaging Telescope (UIT) imaged regions of the LMC in FUV and
NUV bands during the ASTRO-1 flight. Results of these observations, pertaining to
the stellar associations LH52 and LH53, were published by Hill \etal (1995).
The results concerning field stars in the LMC were discussed
in the context of the IMF in the field and in associations (Hill \etal 1994).
Despite the shallowness of the  UIT results, by $\sim$3 mag relative to the HST
data reported here, it is worthwhile first to consider the relative density 
of field stars. We find $\sim$200 stars in 12 arcmin$^2$ to m$_{UV}\approx$19,
while Hill \etal (1994) find that about 70\% of their 1563 stars, to 
m$_{UV}\approx$15, belong to the field population. This yields 
a projected density of 0.9 star arcmin$^{-2}$ for stars brighter than  
m$_{UV}$=16,  from which we expect to find 10 stars brighter than  
m$_{UV}$=16 in our data. Instead, as Fig. 5
demonstrates, we find almost four times this number of stars in our HST
images.

An extensive catalog of UV stars in the two MCs, originating from
both ASTRO-1 (1990) and ASTRO-2 (1995) observations, was published
by Parker \etal (1998). The UIT observations most relevant to the analysis 
presented here are toward N132 and were obtained during the ASTRO-2 flight 
with the B1 filter ($\lambda_e$=1521\AA\,, $\Delta\lambda$=354\AA\,). 
This bandpass is very close to that of the HST WFPC2 F160BW, with a
30\AA\, difference in $\lambda_e$ and the HST bandpass being wider
by 92\AA\ than that of the present observations, therefore a direct
comparison of measured magnitudes is justified. The
LMC area sampled by the HST F160BW mosaic is much smaller than the
UIT image and lies close to the Northern edge of the N132 field.
The list of Parker \etal has been examined and the objects in
common with our data set have been extracted and are presented in Table 4.

A comparison between the UIT B1 magnitudes and those derived here for the
same objects indicates that, with the exception of one object, the UIT
stars are reported brighter by $\sim$1 magnitude than the HST photometry. This 
could, in principle, be the effect of a consistent calibration error of the 
F160BW magnitudes relative to the calibration of UIT, which is derived from
objects in common with IUE. We deem it more likely to be the result of 
confusion contributions in the moderate angular 
resolution UIT images. In fact, in many cases we can identify the stars
which would have been included as single stars in the UIT photometry;
these additional contributors are listed on separate lines in Table 4, below the
primary HST candidate counterpart. Despite the inclusion of the additional
stars, we were not able to reproduce the UIT magnitudes by adding up their
measured flux. Obviously, more objects than the stars we detected must be 
included in order to account for the difference in magnitudes,
from which one can assume that there are even more faint UV stars 
not detected in our WFPC2 images which crowd into the UIT PSF and are
measured as a single star.

Note that the UV survey to be conducted by GALEX (Bianchi \& Martin 1997)
will have similar problems to those of UIT in observing the LMC, as its angular
resolution shall be only 3-5 arcsec. These would only be more acute in
moderately crowded fields, as GALEX is supposed to reach fainter UV
magnitudes than did UIT.

\section{Optical counterparts}

One basic requirement for discussing the nature of the fainter UV sources
is the correspondence with an optical counterpart. Unfortunately,
the F160BW observations in the direction of the LMC were not 
collected together with HST imaging with other filters, thus no images
with comparable resolution in a different spectral band exist in the
HST archives. We attempted photometry on the scans of the Second
STScI Digitized Sky Survey (DSS-II), but the relatively large 
pixel size used for these scans, compounded with the plate scale of
the Schmidt plates, resulted in extreme crowding of the stellar images.
No photometry to a reasonable accuracy could be performed on the stars at the
location of the HST region studied here using the DSS scans.

We searched published catalogs with photometric information and 
 could identify 58 UV sources in Table 3 with objects listed
in the ROE/NRL catalog
(Yentis \etal 1992; Wallin \etal 1994). This catalog is derived from 
COSMOS 5$^{\circ}.4\times$2'.3 scans of the short red UKST survey 
plates in the direction 
of the LMC, which were scanned with a very small laser beam, resulting
in a smaller pixel size than the 15$\mu$m of the DSS-II. The astrometry is
based on the HST Guide Star Catalog and the positions are accurate to
$\sim$2 arcsec. The ROE/NRL catalog also lists magnitudes, which are in a
band close to Johnson R, have a typical accuracy of 0.2 mag, and
reach objects as faint as $\sim$21.5 mag. Our identification of an optical
counterpart for a UV source required a positional coincidence within
2 arcsec; this is justified, given the astrometric accuracy of the 
ROE/NRL catalog, but may result in some chance coincidences. 

The success rate in finding 
optical counterparts by cross-correlating with the ROE/NRL catalog is 78\% 
for objects with m$_{UV}<$16 and drops to $\sim18$\% for m$_{UV}>$18.
The dependence of the success rate on the UV brightness, which should 
in principle be related to the optical brightness of the counterpart,
indicates that most optical identifications are probably not chance
effects. However, the fact that we could not find optical counterparts
for all the bright UV sources is puzzling. One possibility, which we
could not test, is that some possible counterparts were rejected by the
ROE automatic processing of the COSMOS scans because of extreme crowding,
thus they are not listed in the ROE/NRL catalog.

We transformed approximately the ROE/NRL catalog magnitudes to the V band
to determine the nature of the LMC UV sources by assuming that the 
objects are earlier 
than mid-F and converting their UV-R color indices to UV-V. The LMC
objects for which we found optical counterparts in the ROE/NRL 
catalog are listed in Table 5, along with their UV and optical data. The
error in UV--V is obtained as the harmonic mean of the UV photometric
error in Table 3 and an assumed 0.2 mag error in V propagated from the 
``short-red''
magnitude listed in the ROE/NRL catalog and the adopted transformation.

We compare the properties of the LMC stars to stars in the Milky Way
via an observational color-magnitude diagram based on high galactic
latitude objects observed with the FAUST Shuttle-borne telescope. 
We elected to use observed photometry instead of convolving
theoretical model atmosphere spectral energy distributions (SEDs) with
the different instrumental responses, or using IUE spectra ({\it e.g.,} 
Fanelli \etal 1992) convolved similarly, 
because the UV photometric results are unbiased
and UV-selected, just as the LMC sample is unbiased and UV-selected.
The comparison is made with stars observed by the FAUST experiment in
the direction of the North Galactic Pole (NGP), Coma, and Virgo (Brosch
\etal 1995, 1997, 1998) which have parallax measurements in the
Hipparcos and Tycho catalogs. We consider the LMC stars listed in
Table 5 to be at a distance modulus of 18.5 for the purpose of 
determining their absolute magnitudes,
and plot them in the same diagram with the FAUST objects.

The FAUST observations are in a spectral band centered at $\lambda_C$=1650\AA\,
and $\Delta\lambda\sim$300\AA\, wide, which is not too different from the 
WFPC2+F160BW band definition ($\lambda_e$=1491\AA\,, $\Delta\lambda\sim$446\AA\,). 
The two bands mostly overlap, with the HST one being wider and bluer than
that of FAUST. As both bands are not defined extremely accurately, we cannot
derive a color term to transform between the two. It is possible that some
chromatic effects may be present, mainly in the photometry of the 
high T$_{eff}$ stars which are very blue.
We note also that the FAUST observations are biased {\bf against} the detection
of hot main sequence stars because of the direction of observation (high 
$\vert$b$\vert$) and the relative rarity of these stars in the immediate
Solar neighborhood where Hipparcos parallaxes are available. For this 
reason, the earliest main sequence object we can locate on the FAUST 
color-magnitude diagram is late-B.

The UV color-magnitude diagram shown in Fig. 6 indicates that the LMC objects 
join up smoothly with the 
Galactic UV stars in Coma, Virgo, and the North Galactic Pole region. These have 
been identified in the publications mentioned above as main sequence stars. By
inference, we deduce that at least the 58 objects with ROE/NRL catalog
counterparts are also mostly main sequence objects. Note that the
apparent width of the upper main sequence, as defined by the HST data from the
LMC field, is similar to that of the lower main sequence defined by the
FAUST stars in the MW.
 
\section{Discussion}

Figure 5 shows that the F160BW mini-survey of the LMC field becomes 
progressively incomplete
for m$_{UV}>18.5$. Considering the distance modulus, this indicates that we 
are observing LMC objects in the general field with M$_{UV}<0$, but we 
detect objects as
faint as M$_{UV}\simeq$+2.5. The faintest objects for which we have UV-optical
color information may, therefore, be early-F main sequence stars.

We added the UV emission from all the detected stars and found that these
account for 2.53 10$^{-12}$ erg sec$^{-1}$ cm$^{-2}$ \AA\,$^{-1}$ for the
entire area surveyed here. This translates into 
2.49 10$^{-7}$ erg sec$^{-1}$ cm$^{-2}$ \AA\,$^{-1}$ ster$^{-1}$ 
at 1500\AA\, from stellar 
sources not in clusters or associations, and is consistent with the integrated LMC
emission at 1400\AA\, from Page \& Carruthers (1981); their value is higher
by one third, but they include all UV sources, cluster and field, in their estimate.
The average UV brightness measured by TD-1 (Morgan \etal 1979) in the
immediate vicinity of the HST-imaged region of the LMC is 6.8 10$^{-7}$ 
erg sec$^{-1}$ cm$^{-2}$ \AA\,$^{-1}$ ster$^{-1}$ at 1550\AA\,. This is 
higher by a factor of 2.7 than we measure, but the nearest TD-1 measurement
(Region III) lies $\sim$20 arcmin South of the region studied here, deeper 
into the LMC bar.

{\it The calculation shows that the large part of the UV radiation in the LMC, at least
for $\lambda\geq$1500\AA\,, is produced by field stars and not by objects in
clusters or associations.} This is consistent with the estimate by Parker \etal 
(1998) that $\sim$60\% of the UV light in the N132 image originates in
stars fainter than 16 mag or from diffuse emission. It is also
consistent with the measurement
of Meurer \etal (1995) that in starbursts, clusters of young stars provide
at most 20\% of the UV emission.
Our integrated value for the 1500\AA\, emission, combined with the upper limit 
for the diffuse emission at this location and at
2600\AA\, from Vuillemin (1988), indicates a UV color index
[1500]--[2600]$<$0.26 for the integrated stellar population detected in
the UV.


Photometry of field UV sources with the FAUST experiment was done, as
mentioned above,
in a spectral band very similar to that defined by WFPC2 and F160BW. For
comparison purposes, we use here the results for the
NGP (Brosch \etal 1995) and the Virgo regions (Brosch \etal 1997), from 
which we selected only the stellar sources. Data from the Coma field
(Brosch \etal 1998) is not used because this FAUST field is not representative 
of the ``general field'' of the high-latitude MW, as it contains
the Mel 111 open cluster. A comparison with the LMC stars detected
here with the HST is valid, because with the sensitivity limit of FAUST it
observed mostly Galactic sources. At the NGP and Virgo high (Northern) 
Galactic latitude
of the FAUST exposures referred to here we should have a fair indication
of (half) the projected density of UV stars in the Milky Way (MW). In
these regions FAUST detected 172 stars to a UV magnitude limit of
$\sim$14 and a completeness limit of $\sim$12.5.

The FAUST images considered here cover a solid angle
of $\sim$0.05 ster, which projects to $\sim$1.7 10$^4$ pc$^2$ when viewed
from outside the MW and adopting a 1 kpc
scale height for the UV stars. The coverage of the LMC F160BW mosaic is
$\sim$12 square arcmin, which corresponds to $\sim$4400 pc$^2$ at the 
distance of the LMC, thus about one quarter the projected
area in the Milky Way sampled by FAUST. Therefore, the projected number
density of UV stars in the HST observation of the LMC field region
is n$_*\sim\frac{198}{4400}\simeq$0.045 stars pc$^{-2}$, whereas FAUST
counted n$_*\sim\frac{172}{1.7 \, 10^4}\simeq$0.01 stars pc$^{-2}$ in the MW.
The latter value should be doubled to account for the full width of the
Galactic disk when viewed from outside.

It is possible to compare directly the FAUST results with those of the
present HST observations. While FAUST sampled the (thick) Galactic disk and 
the halo, HST observed the edge of the bar and the disk of the LMC
to a similar depth. The reason is that the objects deteced by FAUST
reach into the thick disk and the halo of the MW to $\sim$1 kpc, while
the HST observed objects at $\sim$60 kpc; the gain in sensitivity with HST 
makes up for the increased distance to yield approximately similar depths
of observation.
Thus, if the projected density of UV-bright stars in the LMC and 
in the disk and halo of the MW would have been the same, similar projected 
number densities of stars should have been observed, assuming that
the LMC thickness is probably not much different than that of the MW. 
However, note that the value of n$_*$ for the LMC field is higher than 
that in the MW, as reflected by the FAUST data. The spatial density of 
hot stars in the LMC field regime is apparently higher by a factor of 2-3
 than that in the MW.

{\it We emphasize that our observations indicate a significant contamination
by young field stars in all cluster UV observations in the LMC.} This is especially
important in case one aims to detect blue stragglers, or hot evolved
stars which belong to the target cluster ({\it e.g.,} Mould \etal 1996, 
where four blue stragglers were identified); if the detected objects have
a projected density of $\sim$17 arcmin$^{-2}$ to m$_{UV}\simeq$19 they may
well be only field interlopers, as already remarked by Cole \etal (1997).

Finally, we can use our data, in combination with the similar magnitude
estimates by UIT in N132 (Parker \etal 1998) to produce a combined
(observational) luminosity function for UV stars from m$_{UV}$=8 to 
19. For this purpose we
selected the N132 objects from the LMC catalog of Parker \etal and
scaled our HST stellar UV magnitude distribution to the total number of stars
in the UIT sample with 12$\leq$m$_{UV}\leq$15. This is the 
magnitude range where the
UIT star counts should be essentially complete and where no HST
objects were probably missed. There are 2271 UIT stars
in the N132 region, whereas there are only 17 HST stars in the same 
magnitude range, yielding a scale factor for the number count of 133.59. 
This is close to the ratio of sampled areas between the UIT field and 
that of the HST mosaic ($\sim$105),
providing some confidence for this scaling approach. Note that
the possible $\sim$one magnitude discrepancy between the UIT and HST magnitudes
would not affect our conclusions significantly, because the slope of the 
UV luminosity function is defined mainly by its faint end, {\it i.e.,} by
the HST measurements.

We show in Figure 7 the two luminosity functions plotted on the same scale.
It is evident that the scaling procedure continues the trend of higher 
star counts to m$_{UV}\simeq$18, after which faint stars are progressively
lost from the HST images. Figure 8 shows the logarithm of the combined
star counts, where for m$_{UV}\leq$13 we adoped the UIT star counts and
for the fainter magnitudes those from the (scaled) HST data set. We
compare the slope of this observed luminosity function with the models 
presented in Parker \etal (1998) for the UIT star counts.

The models allow a comparison with an LMC metallicity population formed
continuously over a period $\Delta$t with a Salpeter IMF, using Fig. 11
from Parker \etal and comparing the predicted slope of the star count 
distribution between m$_{UV}$=10 and m$_{UV}$=15 with the same
parameter from the actual, combined star count. The closest fit is with 
a stellar population formed over $\Delta$t$\simeq$1 Gyr, because 
the younger star formation
processes produce a distribution of the fainter UV stars which is much
shallower than observed. This is consistent with the finding of Geha
\etal (1998) that the SF rate in the LMC did not change appreciably
in the last 2 Gyr.

It is similarly possible to compare the slope of predicted
luminosity functions for a continuous star formation over 1 Gyr, with
different IMF slopes (Fig. 14 of Parker \etal), with that  actually
observed. The comparison indicates that the Parker \etal (1998) model 
with the slope most similar to that measured here is that with
an initial mass function slope $\Gamma$ of --1.8, close to the Salpeter value 
and found also to best fit the UIT data. Although the data do not really warrant
it, we note that the measured slope is smaller than that predicted by the
models with $\Gamma$=--1.8 but higher that that for $\Gamma$=--1.0. Thus,
a Salpeter slope with $\Gamma$=--1.4 would probably be better than the
--1.8 value found to best fit the bright end of the UV luminosity
function by Parker \etal 
 
\section{Conclusions}

We analyzed a set of very deep HST images obtained in a UV spectral
band which is not contaminated by visible light leaks. 
Our findings can be summarized as follows:

\begin{enumerate}

\item We identified 198 UV sources with stars in the ``general field'' 
of the LMC, at the edge of the bar and not associated with any known 
star cluster or association. 

\item A comparison with Galactic objects for the 30\% of the
sources for which we found optical catalog counterparts indicates that
the objects are main sequence stars.

\item The UV emission from the stellar sources we detected, scaled to the entire
solid angle covered by the LMC, accounts for 2/3 of the UV emission from
the Cloud.

\item The observed UV luminosity function joins up smoothly with that
defined for brighter objects, but on a wider field, from the UIT observation
of the N 132 region.

\item The joint luminosity function of UV stars, which covers the UV magnitude
range 8 to 19, confirms the claim by Parker 
\etal (1998) that the LMC ``field'' domain has been forming stars 
continuously over (at least) 1 Gyr, with an IMF close to the Salpeter law.

\end{enumerate}
 
\section*{Acknowledgements}
    UV research at Tel Aviv University is supported by grants from
    the Ministry of Science and Arts through the Israel Space Agency,
    from the Austrian Friends of Tel Aviv
    University, and from a Center of Excellence Award from the Israel
    Science Foundation. NB acknowledges support from a US-Israel Binational
    Award to study UV sources measured by the FAUST experiment, and partial 
support from the Space Telescope Science Institute during a sabbatical visit. 
We thank Stefano Casertano for discussions on WFPC2 calibrations, and an
anonymous referee for extensive remarks which improved significantly the
presentation.

\newpage

\section*{References}
\begin{description}

\item Baggett, S., Caserano, S. \& the WFPC2 group 1998, STScI ISR WFPC2 98-01.

\item Battineli, P. \& Demers, S. 1998, AJ, 115, 1472.

\item Bianchi, L. \& Martin, C. 1997 in ``The Ultraviolet 
    Astrophysics beyond
    the IUE final Archive''  (R. Gonzales-Riestra, W. Wamsteker \& R.A. Harris, eds.),
ESA SP-413.

\item Biretta, J. \& Baggett, S. 1998, STScI STAN/WFPC2-no.33, August.

\item Bowyer, S., Sasseen, T.P., Wu, X. \& Lampton, M. 1995, 
ApJS, 96, 461.

\item Brosch, N., Almoznino, E., Leibowitz, E.M.,  Netzer, H., Sasseen, T., 
 Bowyer, S., Lampton, M. \&  Wu  , X.
1995,   ApJ  450, 137.

\item    Brosch, N.,  Formiggini, L., Almoznino, E., Sasseen, T.,
 Lampton, M. \& Bowyer,  S. 
1997,  ApJS,  111, 143.

\item  Brosch, N.,  Ofek, E.O., Almoznino, E.,  Sasseen, T., Lampton, M. \&
 Bowyer, S.
1998, MNRAS, 295, 959.

\item Carruthers, G.R. \& Page, T. 1977, ApJ, 211, 728.

\item Cheng, K.-P., Michalitsianos, A.G., Hintzen, P. Bohlin, R.C., 
O'Connell, R.W., Cornett, R.H., Morton, R.S., Smith, A.M., Smith, E.P. 
\& Stecher, T.P. 1992, ApJ, 395, L29.

\item Clark, D.H., Tuohy, I.R., Dopita, M.A., Mathewson, D.S.,
Long, K.S., Szymkowiak, A.E. \& Culhane, J.L. 1982, ApJ, 225, 440.

\item Cole, A.A., Mould, J.R., Gallagher, J.S. III, Clarke, J.T.,
Trauger, J.T., Ballester, G.E., Burrows, C.J., Casertano, S., Griffiths, R.,
Hester, J.J., Hoessel, J.G. Holtzman, J.A., Scowen, P.A., 
Stapelfeldt, K.R. \& Westphal, J.A. 1997,  AJ, 114, 1945.

\item Court\'{e}s, G., Viton, M., Sivan, J.-P., Decher, R.
\& Gary, A. 1984, Science, 225, 179.

\item Court\'{e}s, G., Viton, M., Bowyer, S., Lampton, M.,
Sasseen, T.P. \& Wu, X. 1995, A\&A, 297, 338.

\item Davies, R.D., Elliott, K.H. \& Meaburn, J. 1976, Mem.
R. astr. Soc., 81, 89.

\item Elson, R.A.W., Gilmore, G.F. \& Santiago, B.X. 1997,
MNRAS, 289, 157.

\item Fanelli, M.N., O'Connell, R.W., Burstein, D.
\& Wu, C.-C. 1992, ApJS, 82, 197.


\item Feast, M.W. 1995 in ``Stellar Populations'' (P.C. van der Kruit
\& G. Gilmore, eds.) Dordrecht; Kluwer Academic Publishers, pp. 153-163.

\item Feitzinger, J.V., Haynes, R.F., Klein, U., Wielebinski, R. \&
Perschke, M. 1987, Vistas in Astr., 30, 243.

\item Geha, M.C., Holtzman, J.A., Mould, J.R., Gallagher, J.S.III,
Watson, A.M., Cole, A.A., Grillmair, C.J., Staplefeld, K.R.,
Ballester, G.E., Burrows, C.J., Clarke, J.T., Crisp, D., Evans, R.W.,
Griffiths, R.E., Hester, J.J., Scowen, P.A., Trauger, J.T. \& Westphal, J.A.
1998, AJ, 115, 1045.

\item Gerola, H. \& Seiden, P.E. 1978, ApJ, 223, 129.

\item Gilmozzi, R., Kinney, E.K., Ewald, S.P., Panagia, N. \& Romaniello, M.
1994, ApJ, 435, L43.

\item Hill, R.S., Isensee, J.E., Cornett, R.H., Bohlin, R.C., O'Connell,
R.W., Roberts, M.S., Smith, A.M. \& Stecher, T.P
1994, ApJ, 425, 122.

\item Hill, R.S., Cheng, K. -P.; Bohlin, R.C., O'Connell, R.W., Roberts, M.S.,
Smith, A.M. \& Stecher, T.P., 1995, ApJ, 446, 622.

\item Holtzman, J.A., Burrows, C.J., Casertano, S., Hester, J.J., 
Trauger, J.T., Watson, A.M. \& Worthey, G. 1995, PASP, 107, 156 (H95a).

\item Holtzman, J.A., Burrows, C.J., Casertano, S., Hester, J.J., 
Trauger, J.T., Watson, A.M. \& Worthey, G. 1995, PASP, 107, 1065 (H95b).

\item Hunter, D.E., Vacca, W.D., Massey, P., Lynds, R. \&
O'Neil, E.J. 1997, AJ, 113, 1691.

\item Kaufman, M. 1979, ApJ, 232, 707.

\item Koornneeff, J. 1977, A\&AS, 29, 117.

\item Leinert, Ch., Bowyer, S., Haikala, L.K., Hanner, M.S.,
Hauser, M.G., Levasseur-Regourd, A.-Ch., Mann, I., Mattila, K.,
Reach, W.T., Schlosser, W., Staude, H.J., Toller, G.N., Weiland, J.L.,
Weinberg, J.L. \& Witt, A.N. 1998, A\&AS, 127, 1.

\item Lucke, P.B. 1974, ApJS, 28, 73.

\item Lucke, P.B.  \& Hodge, P.W. 1971, AJ, 75, 171.

\item MacKenty, J. W., Baggett, S.M., Biretta, J., Hinds, M., 
    Ritchie, C.E., Feinberg, L.D. \& Trauger, J.T. 1995, Proc. SPIE,
    2478, 160.

\item Meurer, G.R., Heckman, T.M., Leitherer, C., Kinney, A., Robert,
C. \& Garnett, D.R. 1995, AJ, 110, 2665.

\item Milliard, B. Donas, J. \& Laget, M. 1991, Adv. Space Res., 11, 135.

\item Morgan, D.H., Nandi, K. \& Carnochan, D.J.  1979, MNRAS, 188, 131.

\item Mould, J.R., Watson, A.M., Gallagher, J.S. III, Ballester, G.E.,
Burrows, C.J., Casertano, S., Clarke, J.T., Crisp, D., Griffiths, R.,
 Hester, J.J., Hoessel, J.G. Holtzman, J.A., Scowen, P.A., 
Stapelfeldt, K.R., Trauger, J.T. \& Westphal, J.A.
1996, ApJ, 461, 762.

\item Nandy, K., Morgan, D.H, \& Carnochan, D.J. 1979, MNRAS, 186, 421.

\item Page, T. \& Carruthers, G.R. 1981, ``Revised listing - S201 
far-ultraviolet atlas of
the Large Magellanic Cloud'', NRL Memorandum Report 4660.

\item Parker, J.Wm., Hill, J.K., Cornett, R.H., Hollis, J., Zamkoff, E.,
Bohlin, R.C., O'Connell, R.W., Neff, S.G., Roberts, M.S.,
Smith, A.M. \& Stecher, T.P. 1998, AJ, 116, 180.

\item Schwering, P.B.W. \& Israel, F.P. 1990 ``Atlas and catalogue
of infrared sources in the Magellanic Clouds'', Dordrecht: Kluwer
Academic Publishers.

\item Smith, A.M., Cornett, R.H. \& Hill, R.S. 1987, ApJ, 320, 609.

\item Stecher, T.P., Baker, G.R. \etal 1992, ApJ, 395, L1.

\item Treyer, M., Ellis, R.S., Milliard, B., Donas, J. \&
Bridges, T.J. 1998, MNRAS, in press (astro-ph/9806056).

\item Thompson, G.I., Nandy, K., Jamar, C., Monfils, A., Houziaux, 
    L., Carnochan, D.J. \& Wilson, R. 1978 ``Catalog of Stellar Ultraviolet 
    Fluxes'', SRC.

\item Vuillemin, A. 1988, A\&AS, 72, 249.

\item  Watson, A.M., Mould, J.R.,
                    Gallagher, J.S.,  Ballester, G.E.,
                    Burrows, C.J., Casertano, S.,
                    Clarke, J.T., Crisp, D., Griffiths, R.E.,
                    Hester, J. J.F, Hoessel, J.G., Holtzman, J.A.,
                    Scowen, P.A., Saplefeldt, K.R.,
                    Trauger, J.T., Westphal, J.A.199, ApJ, 435, 55.

\item Whitmore, B. \& Heyer, I. 1997, Instrument Science Report WFPC2 97-08.

\item Xu, C., Klein, U., Meinert, D., Wielebinski, R. \& Haynes, R.F.
1992, A\&A, 257, 47.

\item Yentis, D.J., Cruddace, R.G., Gursky, H., Stuart, B.V., Wallin, J.F.,
MacGillivray, H.T. \& Collins, C.A. 1992, in ``Digitized optical sky surveys''
(H.T. MacGillivray \& E.B. Thomson, eds.), Dordrecht: Kluwer, p. 67.

\item Wallin, J.F., Yentis, D.J., MacGillivray, H.T., Bauer, S.B. \&
Wong, C.S. 1994, AAS Bull., 184, 2704.

\end{description}

\newpage

\section*{Figure captions}

\figcaption {Mosaic of all LMC F160BW images analyzed here with a few objects
labelled with the numbering in Table 2. We indicate the scale with a 30
arcsec bar near the top right corner and the approximate sky orientation
with directions in the lower right corner. The mosaic is composed of 12
WFPC2 images (see Table 1) and is presented only for general information.
the images were scaled in intensity to approximately match each other
in background level; this was not always fully successful. The salt-and-pepper 
noise is the result of Poisson statistics of the faint UV background
level.}

\figcaption {Number of independent measurements of the same object in
the individual images which make up the mosaic in Fig. 1. The
individual detections are listed in Table 2.}

\figcaption {Single {\it vs.} multiple measurements of the same object. The
histograms represent the distribution of FUV magnitudes for the objects observed
only once (upper panel), and more than once (lower panel). The two
panels have the same X-axis scale. Note the drop-off
in the number counts at m$_{UV}$=19 in both histograms.}

\figcaption {Multiple measurements of the same object {\it vs.} the
adopted average. The solid line represents a perfect slope=1 relation 
between the two independent measurements. The adopted average value is 
represented by empty diamonds.}

\figcaption {Histogram of UV magnitudes for all the objects detected in the HST
observations with F160BW in the LMC.}

\figcaption {Color-magnitude diagram in the UV for stars observed by FAUST 
at high galactic latitude in the Milky Way and by HST in the LMC field (filled
diamonds with error bars). Objects in the NGP field are indicated by  
squares, objects in
Coma by triangles, and objects in Virgo by circles. A few FAUST objects with
lower distance limits (from Hipparcos, with parallax upper limits) are 
indicated by larger symbols with a vertical riser.}

\figcaption {Observed UV luminosity function for the N 132 region.
The distribution was derived from star counts in the UIT frame, to which
we added the star counts from the HST mosaic, scaled by 133.59. The 
counts from the UIT image are represented as filled squares, while those
from the (scaled) HST data are diamonds.}

\figcaption {Combined luminosity function for the UIT and HST data sets.
Note the very good power law approximation for m$_{UV}\geq$14.}


\newpage

\begin{deluxetable}{cccccc}
\tablecaption{WFPC2 datasets analyzed in this paper }
\small
\tablehead{\colhead{Set } & \colhead{Name} &
\colhead{$\alpha$(J2000) } & \colhead{$\delta$(J2000) } & 
\colhead{Exposure (sec.)  }  
& \colhead{ Obs. Date}}
\startdata
A   & u2ou1001t & 5:25:17.6  & --69:30:26       & 2500	   & 230895 \nl
A   & u2ou1101t & 5:25:17.6  & --69:30:26       & 4400     & 230895 \nl
A   & u2ou1201t & 5:25:17.6  & --69:30:26       & 4400     & 230895 \nl
F   & u2ou1301t & 5:25:31.4  & --69:30:52       & 4400     & 230895 \nl
B   & u2ou1501t & 5:25:22.7  & --69:30:09       & 2300     & 260895 \nl
B   & u2ou1502t & 5:25:22.7  & --69:30:09       & 2300     & 260895 \nl
C   & u2ou1601t & 5:25:36.8  & --69:30:30       & 4900     & 260895 \nl
C   & u2ou1701t & 5:25:36.8  & --69:30:30       & 4400     & 260895 \nl
D   & u2ou1e01t & 5:25:17.8  & --69:30:30       & 4600     & 240895 \nl
D   & u2ou1f01t & 5:25:17.8  & --69:30:30       & 4600     & 240895 \nl
E   & u2ou1g01t & 5:25:31.8  & --69:30:51       & 4600     & 240895 \nl
E   & u2ou1h01t & 5:25:31.8  & --69:30:51       & 4600     & 240895 \nl
\enddata

\tablecomments{Exposures with identical pointings are grouped in sets,
which represent essentially CRSPLIT pairs or triplets of images.}

\end{deluxetable}

\newpage
 
\begin{deluxetable}{ccccccc}
\tablecaption{Appendix 1: Far-UV raw photometry of field stars in the LMC }
\small
\tablehead{\colhead{Source no.} & 
\colhead{X} & \colhead{Y} & \colhead{ $\alpha$(J2000)}  
& \colhead{ $\delta$(J2000)} & \colhead{Image \& chip } &
\colhead{FUV-mag (error)}  }
\startdata
1 & 242.117 & 260.523 & 5:25:12.886 & --69:30:37.13 & A-pc & 19.190  (0.041) \nl
1 & 140.889 & 329.237 & 5:25:12.881 & --69:30:37.22 & D-pc & 18.847  (0.035) \nl
2 & 308.600 & 288.237 & 5:25:12.468 & --69:30:39.53 & A-pc & 18.112  (0.020) \nl
2 & 205.689 & 361.042 & 5:25:12.461 & --69:30:39.65 & D-pc & 18.051  (0.021) \nl
3 & 393.000 & 795.000 & 5:25:08.099 & --69:30:35.52 & A-pc & 19.072  (0.034) \nl
4 & 442.279 & 556.773 & 5:25:09.893 & --69:30:41.22 & A-pc & 13.752  (0.002) \nl
4 & 322.418 & 637.817 & 5:25:09.888 & --69:30:41.32 & D-pc & 13.444  (0.002) \nl
5 & 531.523 & 665.313 & 5:25:08.753 & --69:30:43.39 & A-pc & 18.934  (0.033) \nl
5 & 405.079 & 751.607 & 5:25:08.749 & --69:30:43.54 & D-pc & 18.779  (0.032) \nl
6 & 597.824 & 177.715 & 5:25:12.537 & --69:30:53.58 & A-pc & 17.590  (0.010) \nl
6 & 500.647 & 268.901 & 5:25:12.533 & --69:30:53.72 & D-pc & 17.279  (0.013) \nl
7 & 373.000 & 066.000 & 5:25:14.509 & --69:30:50.67 & D-pc & 16.583  (0.009) \nl
8 & 758.794 & 236.243 & 5:25:12.185 & --69:31:05.32 & D-pc & 19.942  (0.081) \nl

10 & 095.365 & 263.299 & 5:25:15.811 & --69:30:07.98 & A-wf2 & 18.220  (0.020) \nl
10 & 106.265 & 306.270 & 5:25:15.812 & --69:30:08.10 & D-wf2 & 18.086  (0.019) \nl
11 & 152.362 & 529.463 & 5:25:16.430 & --69:29:41.14 & A-wf2 & 16.467  (0.008) \nl
11 & 146.375 & 575.363 & 5:25:16.436 & --69:29:41.27 & D-wf2 & 16.221  (0.007) \nl
12 & 182.786 & 067.505 & 5:25:15.895 & --69:29:40.17 & A-wf2 & 19.190  (0.036) \nl
12 & 205.649 & 116.813 & 5:25:13.054 & --69:30:23.56 & D-wf2 & 18.972  (0.034) \nl
13 & 201.310 & 428.790 & 5:25:14.938 & --69:29:49.07 & A-wf2 & 15.717  (0.006) \nl
13 & 201.505 & 477.811 & 5:25:14.953 & --69:29:49.10 & D-wf2 & 15.484  (0.005) \nl
14 & 261.667 & 457.343 & 5:25:14.043 & --69:29:44.40 & A-wf2 & 15.525  (0.005) \nl
14 & 260.222 & 509.980 & 5:25:14.040 & --69:29:44.47 & D-wf2 & 15.318  (0.005) \nl
15 & 292.361 & 174.331 & 5:25:11.744 & --69:30:10.03 & A-wf2 & 19.755  (0.052) \nl
15 & 309.946 & 229.913 & 5:25:11.726 & --69:30:10.14 & D-wf2 & 18.927  (0.032) \nl
16 & 317.258 & 626.340 & 5:25:14.077 & --69:29:26.70 & A-wf2 & 19.116  (0.041) \nl
16 & 305.174 & 682.373 & 5:25:14.078 & --69:29:26.78 & D-wf2 & 18.845  (0.042) \nl
17 & 330.646 & 558.050 & 5:25:13.428 & --69:29:32.66 & A-wf2 & 17.274  (0.013) \nl
17 & 322.711 & 614.787 & 5:25:13.431 & --69:29:32.86 & D-wf2 & 17.088  (0.012) \nl
18 & 336.000 & 789.620 & 5:25:14.726 & --69:29:10.86 & A-wf2 & 17.309  (0.016) \nl
19 & 346.644 & 172.694 & 5:25:10.765 & --69:30:08.47 & A-wf2 & 19.033  (0.032) \nl
19 & 362.435 & 231.432 & 5:25:10.769 & --69:30:08.54 & D-wf2 & 18.731  (0.029) \nl
20 & 531.974 & 152.208 & 5:25:07.318 & --69:30:04.32 & A-wf2 & 18.099  (0.019) \nl
20 & 548.410 & 222.324 & 5:25:07.333 & --69:30:04.46 & D-wf2 & 17.855  (0.017) \nl
21 & 714.051 & 479.000 & 5:25:06.085 & --69:29:27.74 & A-wf2 & 17.795  (0.021) \nl
21 & 710.531 & 559.679 & 5:25:06.095 & --69:29:27.93 & D-wf2 V & 17.718  (0.030) \nl
22 & 719.335 & 417.689 & 5:25:05.614 & --69:29:33.39 & A-wf2 & 21.144  (0.282) \nl
22 & 729.462 & 514.000 & 5:25:05.523 & --69:29:31.72 & D-wf2 & 17.370  (0.021) \nl
23 & 750.094 & 434.000 & 5:25:05.170 & --69:29:30.81 & A-wf2 & 19.497  (0.084) \nl
23 & 716.000 & 499.988 & 5:25:05.687 & --69:29:33.40 & D-wf2 & 19.820  (0.134) \nl
24 & 089.643 & 459.087 & 5:25:17.123 & --69:29:49.75 & A-wf2 & 19.532  (0.045) \nl
24 & 088.011 & 501.180 & 5:25:17.119 & --69:29:49.89 & D-wf2 & 19.260  (0.039) \nl
25 & 645.560 & 028.878 & 5:25:04.619 & --69:30:20.24 & D-wf2 & 15.771  (0.006) \nl
26 & 771.340 & 076.698 & 5:25:02.585 & --69:30:12.39 & D-wf2 & 17.360  (0.015) \nl

31  & 054.017 & 397.253 & 5:25:21.550 & --69:30:41.15 & A-wf3 & 17.643  (0.015) \nl
31  & 072.656 & 372.330 & 5:25:21.545 & --69:30:41.39 & D-wf3 & 17.381  (0.013) \nl
31  & 178.305 & 274.530 & 5:25:21.556 & --69:30:40.94 & B-wf4 & 17.330  (0.015) \nl

32  & 067.471 & 563.700 & 5:25:24.590 & --69:30:45.32 & A-wf3 & 15.895  (0.006) \nl
32  & 075.996 & 539.289 & 5:25:24.602 & --69:30:45.46 & D-wf3 & 15.672  (0.006) \nl
32  & 344.651 & 269.488 & 5:25:24.597 & --69:30:45.08 & B-wf4 & 15.584  (0.006) \nl 

33  & 091.434 & 478.313 & 5:25:23.224 & --69:30:40.35 & A-wf3 & 18.351  (0.022) \nl
33  & 104.822 & 455.424 & 5:25:23.223 & --69:30:40.48 & D-wf3 & 18.103  (0.020) \nl
33  & 260.864 & 241.980 & 5:25:23.220 & --69:30:40.07 & B-wf4 & 17.907  (0.020) \nl

34  & 275.671 & 633.669 & 5:25:27.136 & --69:30:28.18 & A-wf3 & 18.748  (0.033) \nl
34  & 279.313 & 621.811 & 5:25:27.154 & --69:30:28.35 & D-wf3 & 18.488  (0.030) \nl
34  & 425.554 & 065.323 & 5:25:27.152 & --69:30:27.96 & B-wf4 & 18.497  (0.032) \nl

35  & 303.764 & 447.109 & 5:25:23.991 & --69:30:19.50 & A-wf3 & 18.142  (0.020) \nl
35  & 318.739 & 437.390 & 5:25:23.987 & --69:30:19.67 & D-wf3 & 17.937  (0.019) \nl
120 & 043.898 & 242.704 & 5:25:23.966 & --69:30:19.70 & B-wf3 & 17.988  (0.024) \nl

36  & 361.684 & 320.639 & 5:25:22.082 & --69:30:09.89 & A-wf3 & 18.525  (0.025) \nl
36  & 384.462 & 314.714 & 5:25:22.079 & --69:30:10.12 & D-wf3 & 18.351  (0.024) \nl
121 & 109.647 & 119.213 & 5:25:22.071 & --69:30:10.18 & B-wf3 & 18.411  (0.029) \nl
 
37a & 453.913 & 587.461 & 5:25:27.433 & --69:30:09.89 & A-wf3 & 19.250  (0.054) \nl
37a & 459.992 & 586.638 & 5:25:27.429 & --69:30:10.07 & D-wf3 & 18.830  (0.044) \nl
122 & 187.228 & 392.523 & 5:25:27.417 & --69:30:10.17 & B-wf3 & 19.176  (0.054) \nl

37b & 457.481 & 572.207 & 5:25:27.185 & --69:30:09.12 & A-wf3 & 21.788  (0.498) \nl
37b & 463.361 & 566.679 & 5:25:27.081 & --69:30:09.24 & D-wf3 & 20.348  (0.153) \nl
38a & 466.876 & 556.314 & 5:25:26.955 & --69:30:07.75 & A-wf3 & 19.474  (0.064) \nl
38a & 474.728 & 557.458 & 5:25:26.974 & --69:30:07.95 & D-wf3 & 19.555  (0.077) \nl
38b & 471.438 & 549.874 & 5:25:26.862 & --69:30:07.05 & A-wf3 & 20.002  (0.098) \nl
38b & 483.000 & 551.000 & 5:25:26.911 & --69:30:06.92 & D-wf3 & 21.066  (0.297) \nl
39  & 535.855 & 227.626 & 5:25:21.510 & --69:29:50.51 & A-wf3 & 19.375  (0.044) \nl
39  & 563.767 & 232.487 & 5:25:21.514 & --69:29:50.69 & D-wf3 & 19.160  (0.044) \nl

40  & 551.226 & 293.726 & 5:25:22.788 & --69:29:51.16 & A-wf3 & 19.865  (0.066) \nl
40  & 574.895 & 299.707 & 5:25:22.813 & --69:29:51.36 & D-wf3 & 19.649  (0.066) \nl
126 & 301.585 & 104.793 & 5:25:22.770 & --69:29:51.48 & B-wf3 & 19.584  (0.069) \nl

41  & 591.000 & 588.000 & 5:25:28.300 & --69:29:57.07 & A-wf3 & 18.666  (0.042) \nl
41  & 594.612 & 595.292 & 5:25:28.275 & --69:29:57.50 & D-wf3 & 20.071  (0.159) \nl
42  & 612.000 & 447.605 & 5:25:25.918 & --69:29:50.48 & A-wf3 & 17.370  (0.015) \nl
42  & 622.056 & 472.000 & 5:25:26.184 & --69:29:51.50 & D-wf3 & 18.517  (0.036) \nl
43  & 612.000 & 379.000 & 5:25:24.704 & --69:29:48.25 & A-wf3 & 17.819  (0.018) \nl
43  & 639.653 & 383.644 & 5:25:24.652 & --69:29:47.48 & D-wf3 & 18.228  (0.028) \nl

44  & 607.375 & 335.463 & 5:25:23.887 & --69:29:47.28 & A-wf3 & 17.379  (0.014) \nl
44  & 628.585 & 344.541 & 5:25:23.886 & --69:29:47.48 & D-wf3 & 17.239  (0.014) \nl
127 & 355.420 & 150.181 & 5:25:23.879 & --69:29:47.56 & B-wf3 & 17.319  (0.014) \nl

45  & 708.733 & 534.000 & 5:25:28.060 & --69:29:44.36 & A-wf3 & 18.843  (0.067) \nl
45  & 713.612 & 546.306 & 5:25:27.988 & --69:29:44.84 & D-wf3 V & 21.214  (0.713) \nl
46  & 793.000 & 300.134 & 5:25:24.429 & --69:29:28.83 & A-wf3 & 18.695  (0.042) \nl
51 & 047.224 & 456.328 & 5:25:12.589 & --69:31:10.06 & A-wf4 & 18.857  (0.064) \nl
52 & 084.929 & 118.391 & 5:25:15.438 & --69:30:39.84 & A-wf4 & 19.605  (0.098) \nl
52 & 053.441 & 080.329 & 5:25:15.422 & --69:30:39.98 & D-wf4 & 18.972  (0.041) \nl

53 & 094.265 & 149.601 & 5:25:15.396 & --69:30:43.01 & A-wf4 & 14.354  (0.008) \nl
53 & 060.674 & 112.426 & 5:25:15.375 & --69:30:43.18 & D-wf4 & 14.151  (0.003) \nl
104 & 792.612 & 429.584 & 5:25:15.426 & --69:30:43.56 & B-pc & 14.212  (0.003) \nl

54 & 098.477 & 754.720 & 5:25:11.608 & --69:31:39.53 & A-wf4 & 19.469  (0.097) \nl
55 & 109.924 & 713.486 & 5:25:12.074 & --69:31:36.14 & A-wf4 & 18.812  (0.067) \nl
56 & 135.128 & 552.157 & 5:25:13.531 & --69:31:21.99 & A-wf4 & 16.911  (0.026) \nl
56 & 077.397 & 516.988 & 5:25:13.519 & --69:31:22.15 & D-wf4 & 16.939  (0.011) \nl
57 & 143.718 & 188.409 & 5:25:16.011 & --69:30:48.25 & A-wf4 & 17.467  (0.034) \nl
57 & 108.329 & 153.795 & 5:25:16.019 & --69:30:48.35 & D-wf4 & 17.254  (0.013) \nl
58 & 206.444 & 734.497 & 5:25:13.637 & --69:31:41.34 & A-wf4 & 19.138  (0.080) \nl
58 & 159.454 & 706.861 & 5:25:14.009 & --69:31:42.41 & D-wf4 & 20.012  (0.093) \nl
59 & 206.333 & 553.424 & 5:25:14.783 & --69:31:24.48 & A-wf4 & 19.166  (0.078) \nl
59 & 148.403 & 522.668 & 5:25:14.773 & --69:31:24.61 & D-wf4 & 18.894  (0.032) \nl

60 & 210.734 & 485.336 & 5:25:15.289 & --69:31:18.25 & A-wf4 & 18.633  (0.059) \nl
60 & 157.354 & 454.904 & 5:25:15.296 & --69:31:18.37 & D-wf4 & 18.395  (0.024) \nl
61 & 227.305 & 219.682 & 5:25:17.303 & --69:30:53.94 & A-wf4 & 14.985  (0.011) \nl
61 & 189.720 & 190.583 & 5:25:17.285 & --69:30:54.08 & D-wf4 & 14.857  (0.004) \nl
62 & 253.328 & 175.909 & 5:25:18.043 & --69:30:50.80 & A-wf4 & 17.814  (0.040) \nl
62 & 218.465 & 148.465 & 5:25:18.035 & --69:30:50.90 & D-wf4 & 17.585  (0.015) \nl
63 & 257.563 & 532.684 & 5:25:15.824 & --69:31:24.24 & A-wf4 & 18.909  (0.068) \nl
63 & 200.553 & 503.562 & 5:25:15.815 & --69:31:24.25 & D-wf4 & 19.600  (0.054) \nl
64 & 342.984 & 796.393 & 5:25:15.676 & --69:31:51.71 & A-wf4 & 16.687  (0.024) \nl
64 & 269.822 & 773.505 & 5:25:15.667 & --69:31:51.87 & D-wf4 & 16.149  (0.008) \nl
65 & 367.672 & 614.385 & 5:25:17.258 & --69:31:35.62 & A-wf4 & 18.945  (0.072) \nl
65 & 305.810 & 593.494 & 5:25:17.244 & --69:31:35.77 & D-wf4 & 18.705  (0.032) \nl
66 & 332.477 & 179.500 & 5:25:19.430 & --69:30:53.73 & A-wf4 & 19.378  (0.086) \nl
66 & 297.485 & 156.963 & 5:25:19.421 & --69:30:53.93 & D-wf4 & 19.060  (0.036) \nl
67 & 380.004 & 751.588 & 5:25:16.617 & --69:31:48.79 & A-wf4 & 18.907  (0.074) \nl
67 & 309.666 & 731.261 & 5:25:16.591 & --69:31:48.99 & D-wf4 & 18.739  (0.039) \nl
68 & 458.448 & 160.663 & 5:25:21.800 & --69:30:56.21 & A-wf4 & 20.619  (0.169) \nl
68 & 424.491 & 145.929 & 5:25:21.791 & --69:30:56.33 & D-wf4 & 20.102  (0.080) \nl
69 & 473.572 & 554.724 & 5:25:19.533 & --69:31:33.58 & A-wf4 & 19.108  (0.078) \nl
69 & 413.697 & 538.809 & 5:25:19.500 & --69:31:33.55 & D-wf4 & 19.334  (0.050) \nl
70 & 499.100 & 538.825 & 5:25:20.100 & --69:31:32.96 & A-wf4 & 20.228  (0.151) \nl
70 & 441.542 & 526.181 & 5:25:20.074 & --69:31:33.18 & D-wf4 & 19.673  (0.068) \nl
71 & 483.147 & 229.535 & 5:25:21.800 & --69:31:03.49 & A-wf4 & 18.764  (0.064) \nl
71 & 444.793 & 216.463 & 5:25:21.779 & --69:31:03.65 & D-wf4 & 18.507  (0.026) \nl
72 & 575.000 & 691.462 & 5:25:20.468 & --69:31:49.74 & A-wf4 & 18.744  (0.076) \nl
72 & 507.595 & 682.304 & 5:25:20.444 & --69:31:49.85 & D-wf4 & 18.388  (0.034) \nl
73 & 637.564 & 553.584 & 5:25:22.458 & --69:31:38.97 & A-wf4 & 16.397  (0.021) \nl
73 & 578.677 & 549.316 & 5:25:22.442 & --69:31:39.18 & D-wf4 & 16.211  (0.008) \nl
74 & 661.381 & 711.638 & 5:25:21.864 & --69:31:54.43 & A-wf4 & 23.046  (5.329) \nl
75 & 690.004 & 709.468 & 5:25:22.390 & --69:31:55.20 & A-wf4 V & 16.453  (0.027)  \nl
75 & 629.224 & 724.146 & 5:25:22.429 & --69:31:57.16 & D-wf4 V & 16.275  (0.012) \nl
76 & 735.370 & 686.709 & 5:25:23.333 & --69:31:54.54 & A-wf4 V & 15.700  (0.019) \nl
76 & 668.214 & 687.938 & 5:25:23.332 & --69:31:54.73 & D-wf4 V & 15.748  (0.008) \nl
77 & 798.535 & 543.642 & 5:25:25.366 & --69:31:43.37 & A-wf4 V & INDEF  INDEF \nl
77 & 739.637 & 549.094 & 5:25:25.352 & --69:31:43.60 & D-wf4 V & 14.723  (0.004) \nl
78 & 655.113 & 651.461 & 5:25:22.145 & --69:31:48.68 & A-wf4 & 16.992  (0.029) \nl
78 & 590.229 & 647.508 & 5:25:22.136 & --69:31:48.82 & D-wf4 & 16.882  (0.013) \nl
79 & 759.209 & 568.999 & 5:25:25.604 & --69:31:46.04 & D-wf4 V & 16.451  (0.014) \nl
80 & 760.697 & 349.908 & 5:25:26.789 & --69:31:25.25 & D-wf4 & 15.582  (0.006) \nl
81 & 773.092 & 177.880 & 5:25:27.921 & --69:31:09.27 & D-wf4 & 19.485  (0.080) \nl
82 & 673.152 & 602.838 & 5:25:22.780 & --69:31:44.73 & A-wf4 & 19.070  (0.094) \nl
82 & 608.571 & 600.550 & 5:25:22.713 & --69:31:44.86 & D-wf4 & 18.372  (0.036) \nl
83 & 627.839 & 723.154 & 5:25:21.202 & --69:31:54.46 & A-wf4 V & 19.087  (0.134)  \nl
83 & 556.981 & 718.108 & 5:25:21.159 & --69:31:54.63 & D-wf4 & 19.264  (0.094) \nl

100 & 168.000 & 446.000 & 5:25:16.766 & --69:30:16.13 & B-pc & 18.408  (0.028) \nl
101 & 215.000 & 513.000 & 5:25:16.097 & --69:30:17.35 & B-pc & 19.122  (0.051) \nl
102 & 635.401 & 631.425 & 5:25:14.121 & --69:30:34.21 & B-pc & 18.717  (0.037) \nl
103 & 647.000 & 632.000 & 5:25:14.085 & --69:30:34.72 & B-pc & 18.640  (0.036) \nl

110 & 299.000 & 701.661 & 5:25:19.213 & --69:29:04.33 & B-wf2 & 17.863  (0.021) \nl
111 & 300.088 & 387.000 & 5:25:17.617 & --69:29:34.38 & B-wf2 & 18.471  (0.028) \nl
112 & 297.306 & 361.398 & 5:25:17.540 & --69:29:36.95 & B-wf2 & 19.023  (0.041) \nl
113 & 347.844 & 719.000 & 5:25:18.430 & --69:29:01.34 & B-wf2 & 18.084  (0.025) \nl
114 & 406.051 & 641.000 & 5:25:16.966 & --69:29:07.22 & B-wf2 & 19.030  (0.050) \nl
115 & 483.812 & 398.539 & 5:25:14.333 & --69:29:28.46 & B-wf2 & 18.852  (0.036) \nl
116 & 622.497 & 323.459 & 5:25:11.424 & --69:29:31.95 & B-wf2 & 18.583  (0.033) \nl
117 & 745.000 & 439.029 & 5:25:09.786 & --69:29:17.60 & B-wf2 & 19.280  (0.077) \nl

123 & 225.288 & 772.000 & 5:25:34.508 & --69:30:16.64 & B-wf3 & 17.930  (0.035) \nl
124 & 241.545 & 725.462 & 5:25:33.743 & --69:30:13.89 & B-wf3 & 18.377  (0.048) \nl
125 & 303.973 & 496.000 & 5:25:29.912 & --69:30:01.78 & B-wf3 & 19.302  (0.065) \nl 

128 & 378.000 & 650.865 & 5:25:33.092 & --69:29:58.83 & B-wf3 & 18.508  (0.052) \nl
129 & 412.622 & 530.883 & 5:25:31.087 & --69:29:52.35 & B-wf3 & 18.764  (0.050) \nl
130 & 529.365 & 075.459 & 5:25:23.422 & --69:29:28.92 & B-wf3 & 19.852  (0.090) \nl
131 & 584.646 & 510.429 & 5:25:31.605 & --69:29:35.35 & B-wf3 & 17.379  (0.022) \nl
132 & 653.131 & 348.662 & 5:25:29.013 & --69:29:24.40 & B-wf3 & 18.302  (0.041) \nl 
133 & 707.435 & 485.697 & 5:25:31.777 & --69:29:22.96 & B-wf3 V & 17.089  (0.021) \nl
134 & 771.515 & 521.146 & 5:25:32.752 & --69:29:17.87 & B-wf3 V & 19.138  (0.256) \nl
135 & 786.484 & 300.654 & 5:25:28.825 & --69:29:10.50 & B-wf3 & 17.318  (0.025) \nl

140 & 105.858 & 432.353 & 5:25:19.410 & --69:30:54.01 & B-wf4 & 19.175  (0.050) \nl
141 & 253.847 & 491.243 & 5:25:21.776 & --69:31:03.78 & B-wf4 & 18.505  (0.032) \nl

142 & 488.014 & 783.042 & 5:25:24.477 & --69:31:38.04 & B-wf4 & 18.736  (0.078) \nl
143 & 568.832 & 625.478 & 5:25:26.787 & --69:31:25.34 & B-wf4 & 15.517  (0.006) \nl
144 & 601.438 & 272.557 & 5:25:29.249 & --69:30:52.58 & B-wf4 & 19.118  (0.055) \nl
145 & 612.297 & 329.781 & 5:25:29.146 & --69:30:58.32 & B-wf4 & 18.940  (0.057) \nl
146 & 629.694 & 328.539 & 5:25:29.459 & --69:30:58.70 & B-wf4 & 17.674  (0.022) \nl
147 & 704.360 & 415.911 & 5:25:30.346 & --69:31:09.18 & B-wf4 & 19.057  (0.081) \nl
148 & 729.542 & 595.259 & 5:25:29.839 & --69:31:26.88 & B-wf4 V & 15.263  (0.007) \nl
149 & 749.617 & 198.932 & 5:25:32.299 & --69:30:49.79 & B-wf4 & 17.924  (0.032) \nl

150 & 377.802 & 209.461 & 5:25:32.266 & --69:30:50.04 & C-pc & 17.332  (0.018) \nl
151 & 657.185 & 470.329 & 5:25:29.433 & --69:30:58.98 & C-pc & 16.819  (0.012) \nl

160 & 068.322 & 145.245 & 5:25:34.631 & --69:30:25.36 & C-wf2 & 18.601  (0.039) \nl
161 & 088.595 & 469.467 & 5:25:35.909 & --69:29:53.87 & C-wf2 & 17.623  (0.018) \nl
162 & 093.028 & 072.716 & 5:25:33.807 & --69:30:31.64 & C-wf2 & 19.435  (0.073) \nl
163 & 144.503 & 243.762 & 5:25:33.751 & --69:30:14.05 & C-wf2 & 17.623  (0.018) \nl
164 & 142.619 & 086.538 & 5:25:32.990 & --69:30:29.06 & C-wf2 & 19.019  (0.052) \nl
165 & 356.613 & 587.368 & 5:25:31.621 & --69:29:35.47 & C-wf2 & 16.552  (0.011) \nl
166 & 380.868 & 710.416 & 5:25:31.780 & --69:29:23.07 & C-wf2 & 16.193  (0.010) \nl
167 & 476.880 & 190.383 & 5:25:27.413 & --69:30:10.35 & C-wf2 & 18.190  (0.028) \nl
168 & 499.370 & 213.699 & 5:25:27.127 & --69:30:07.56 & C-wf2 & 18.937  (0.049) \nl
169 & 506.226 & 205.546 & 5:25:26.956 & --69:30:08.14 & C-wf2 & 18.945  (0.049) \nl
170 & 475.997 & 458.809 & 5:25:28.786 & --69:29:44.57 & C-wf2 & 18.148  (0.031) \nl
171 & 517.928 & 656.790 & 5:25:29.015 & --69:29:24.61 & C-wf2 & 17.389  (0.027) \nl
172 & 628.318 & 047.514 & 5:25:23.963 & --69:30:19.93 & C-wf2 & 17.052  (0.014) \nl
173 & 718.884 & 359.675 & 5:25:23.866 & --69:29:47.74 & C-wf2 & 16.350  (0.011) \nl
174 & 751.491 & 113.581 & 5:25:22.071 & --69:30:10.36 & C-wf2 & 17.248  (0.019) \nl

180 & 051.431 & 142.692 & 5:25:36.224 & --69:30:38.05 & C-wf3 & 18.946  (0.060) \nl
181 & 065.528 & 505.268 & 5:25:42.880 & --69:30:46.50 & C-wf3 & 17.114  (0.014) \nl
182 & 072.679 & 141.546 & 5:25:36.310 & --69:30:36.03 & C-wf3 & 18.104  (0.028) \nl
183 & 114.264 & 105.729 & 5:25:35.871 & --69:30:31.07 & C-wf3 & 18.040  (0.027) \nl
184 & 148.323 & 181.513 & 5:25:37.413 & --69:30:29.90 & C-wf3 & 18.058  (0.028) \nl
185 & 202.542 & 443.744 & 5:25:42.451 & --69:30:31.82 & C-wf3 & 17.172  (0.015) \nl
186 & 257.723 & 099.560 & 5:25:36.480 & --69:30:17.29 & C-wf3 & 19.778  (0.109) \nl
187 & 281.897 & 420.586 & 5:25:42.442 & --69:30:23.55 & C-wf3 & 17.904  (0.028) \nl
188 & 296.206 & 425.715 & 5:25:42.605 & --69:30:22.35 & C-wf3 & 16.062  (0.008) \nl
189 & 353.159 & 510.794 & 5:25:44.447 & --69:30:19.17 & C-wf3 & 13.448  (0.002) \nl
200 & 374.164 & 500.066 & 5:25:44.372 & --69:30:16.89 & C-wf3 & 17.764  (0.026) \nl
201 & 397.615 & 345.481 & 5:25:41.664 & --69:30:10.52 & C-wf3 & 17.503  (0.019) \nl
202 & 424.650 & 161.750 & 5:25:38.456 & --69:30:02.98 & C-wf3 & 18.740  (0.044) \nl
203 & 535.846 & 358.931 & 5:25:42.610 & --69:29:57.64 & C-wf3 & 13.922  (0.003) \nl
204 & 570.584 & 293.642 & 5:25:41.606 & --69:29:52.55 & C-wf3 & 17.414  (0.019) \nl
205 & 570.775 & 210.666 & 5:25:40.098 & --69:29:50.34 & C-wf3 & 17.940  (0.027) \nl
206 & 584.649 & 525.937 & 5:25:45.901 & --69:29:57.45 & C-wf3 & 16.287  (0.011) \nl
207 & 668.161 & 219.805 & 5:25:40.767 & --69:29:41.24 & C-wf3 & 18.819  (0.063) \nl

210 & 083.143 & 175.582 & 5:25:34.401 & --69:30:50.67 & C-wf4 & 17.310  (0.016) \nl
211 & 089.948 & 147.000 & 5:25:34.660 & --69:30:48.19 & C-wf4 & 18.478  (0.037) \nl
212 & 117.067 & 153.880 & 5:25:35.128 & --69:30:49.61 & C-wf4 & 19.662  (0.095) \nl
213 & 130.968 & 461.566 & 5:25:33.725 & --69:31:19.20 & C-wf4 & 15.590  (0.006) \nl
214 & 132.439 & 438.356 & 5:25:33.882 & --69:31:17.06 & C-wf4 & 17.674  (0.022) \nl
215 & 173.487 & 145.009 & 5:25:36.188 & --69:30:50.29 & C-wf4 & 15.173  (0.005) \nl
216 & 235.989 & 625.341 & 5:25:34.780 & --69:31:37.80 & C-wf4 & 18.937  (0.059) \nl
217 & 279.585 & 274.929 & 5:25:37.422 & --69:31:05.48 & C-wf4 & 17.918  (0.025) \nl
218 & 319.086 & 620.000 & 5:25:36.312 & --69:31:39.65 & C-wf4 & 18.998  (0.064) \nl
219 & 397.579 & 230.891 & 5:25:39.800 & --69:31:04.67 & C-wf4 & 16.305  (0.009) \nl
220 & 423.869 & 126.613 & 5:25:40.826 & --69:30:55.40 & C-wf4 & 14.387  (0.003) \nl
221 & 470.121 & 563.511 & 5:25:39.361 & --69:31:38.42 & C-wf4 & 13.829  (0.002) \nl
222 & 508.547 & 240.440 & 5:25:41.768 & --69:31:08.63 & C-wf4 & 16.744  (0.011) \nl
223 & 581.176 & 315.656 & 5:25:42.697 & --69:31:17.83 & C-wf4 & 18.833  (0.055) \nl
224 & 600.979 & 310.116 & 5:25:43.086 & --69:31:17.91 & C-wf4 & 19.132  (0.073) \nl
225 & 668.361 & 305.426 & 5:25:44.324 & --69:31:19.30 & C-wf4 & 18.497  (0.047) \nl
226 & 732.990 & 129.566 & 5:25:46.412 & --69:31:04.39 & C-wf4 & 15.352  (0.006) \nl
227 & 747.563 & 627.673 & 5:25:44.034 & --69:31:52.10 & C-wf4 V & 16.911  (0.051) \nl
228 & 758.000 & 500.755 & 5:25:44.912 & --69:31:40.36 & C-wf4 V & 18.298  (0.086) \nl
229 & 731.764 & 296.000 & 5:25:45.507 & --69:31:20.20 & C-wf4 & 15.889  (0.008) \nl

230 & 058.090 & 084.796 & 5:25:29.128 & --69:30:58.46 & E-pc & 18.231  (0.042) \nl
231 & 706.834 & 185.571 & 5:25:26.768 & --69:31:25.43 & E-pc & 14.984  (0.004) \nl

240 & 072.732 & 070.708 & 5:25:29.258 & --69:30:53.16 & E-wf2 & 18.284  (0.031) \nl
241 & 156.000 & 326.598 & 5:25:29.033 & --69:30:26.59 & E-wf2 & 18.956  (0.048) \nl
242 & 139.914 & 688.479 & 5:25:31.124 & --69:29:52.35 & E-wf2 & 18.108  (0.031) \nl
243 & 237.294 & 301.000 & 5:25:27.431 & --69:30:26.86 & E-wf2 & 17.470  (0.017) \nl
244 & 247.279 & 281.896 & 5:25:27.153 & --69:30:28.42 & E-wf2 & 17.685  (0.020) \nl
245 & 279.606 & 732.290 & 5:25:28.819 & --69:29:44.47 & E-wf2 & 18.230  (0.046) \nl
246 & 281.311 & 462.645 & 5:25:27.439 & --69:30:10.25 & E-wf2 & 18.078  (0.026) \nl
247 & 300.837 & 466.093 & 5:25:27.112 & --69:30:09.36 & E-wf2 & 19.797  (0.106) \nl
248 & 270.900 & 207.564 & 5:25:26.336 & --69:30:34.98 & E-wf2 & 19.185  (0.060) \nl
249 & 331.142 & 079.421 & 5:25:24.598 & --69:30:45.61 & E-wf2 & 14.776  (0.004) \nl
250 & 379.159 & 474.646 & 5:25:25.710 & --69:30:06.51 & E-wf2 & 20.190  (0.149) \nl
251 & 392.711 & 475.000 & 5:25:25.459 & --69:30:06.04 & E-wf2 & 19.592  (0.086) \nl
252 & 431.162 & 322.179 & 5:25:23.994 & --69:30:19.73 & E-wf2 & 17.086  (0.013) \nl
253 & 461.645 & 303.259 & 5:25:23.350 & --69:30:20.76 & E-wf2 & 19.084  (0.053) \nl
254 & 414.554 & 108.455 & 5:25:23.230 & --69:30:40.66 & E-wf2 & 17.120  (0.013) \nl
255 & 498.102 & 076.538 & 5:25:21.541 & --69:30:41.48 & E-wf2 & 16.527  (0.009) \nl
256 & 522.289 & 632.635 & 5:25:23.896 & --69:29:47.61 & E-wf2 & 16.285  (0.011) \nl
257 & 553.549 & 388.252 & 5:25:22.100 & --69:30:10.17 & E-wf2 & 17.394  (0.018) \nl
258 & 635.320 & 568.437 & 5:25:21.518 & --69:29:50.78 & E-wf2 & 18.085  (0.049) \nl
259 & 662.106 & 702.000 & 5:25:21.709 & --69:29:37.33 & E-wf2 V & 16.176  (0.036) \nl
260 & 668.000 & 759.000 & 5:25:21.890 & --69:29:31.79 & E-wf2 V & 16.881  (0.192) \nl

270 & 054.313 & 503.182 & 5:25:37.873 & --69:31:08.27 & E-wf3 & 17.795  (0.023) \nl
271 & 071.041 & 473.684 & 5:25:37.414 & --69:31:05.86 & E-wf3 & 18.086  (0.028) \nl
272 & 069.799 & 675.279 & 5:25:41.076 & --69:31:11.26 & E-wf3 & 18.917  (0.067) \nl
273 & 079.882 & 699.812 & 5:25:41.579 & --69:31:10.96 & E-wf3 & 17.604  (0.025) \nl
274 & 102.244 & 704.530 & 5:25:41.765 & --69:31:08.99 & E-wf3 & 16.773  (0.013) \nl
275 & 113.652 & 592.643 & 5:25:39.793 & --69:31:05.02 & E-wf3 & 16.423  (0.009) \nl 
276 & 149.598 & 165.465 & 5:25:32.211 & --69:30:50.16 & E-wf3 & 17.467  (0.018) \nl
277 & 172.319 & 279.436 & 5:25:34.395 & --69:30:51.05 & E-wf3 & 17.335  (0.016) \nl
278 & 202.425 & 369.526 & 5:25:36.187 & --69:30:50.61 & E-wf3 & 15.284  (0.005) \nl
279 & 217.944 & 620.025 & 5:25:40.840 & --69:30:55.78 & E-wf3 & 14.398  (0.003) \nl
280 & 251.000 & 728.724 & 5:25:42.966 & --69:30:55.47 & E-wf3 & 17.932  (0.041) \nl
281 & 338.737 & 699.430 & 5:25:42.888 & --69:30:46.41 & E-wf3 & 16.822  (0.016) \nl
282 & 325.910 & 472.064 & 5:25:38.700 & --69:30:41.51 & E-wf3 & 18.868  (0.054) \nl
283 & 308.327 & 426.849 & 5:25:37.768 & --69:30:42.00 & E-wf3 & 16.365  (0.009) \nl
284 & 326.544 & 338.481 & 5:25:36.255 & --69:30:37.91 & E-wf3 & 18.975  (0.054) \nl
285 & 347.444 & 336.793 & 5:25:36.326 & --69:30:35.85 & E-wf3 & 17.933  (0.025) \nl
286 & 311.361 & 210.011 & 5:25:33.848 & --69:30:35.91 & E-wf3 & 19.317  (0.074) \nl
287 & 388.582 & 301.690 & 5:25:35.898 & --69:30:30.98 & E-wf3 & 17.887  (0.025) \nl
288 & 421.748 & 376.416 & 5:25:37.435 & --69:30:29.83 & E-wf3 & 17.912  (0.024) \nl
289 & 426.167 & 217.360 & 5:25:34.564 & --69:30:25.08 & E-wf3 & 18.697  (0.044) \nl
290 & 475.195 & 638.307 & 5:25:42.482 & --69:30:31.68 & E-wf3 & 16.914  (0.019) \nl
291 & 523.612 & 141.245 & 5:25:33.684 & --69:30:13.77 & E-wf3 & 17.822  (0.023) \nl 
292 & 568.756 & 620.401 & 5:25:42.627 & --69:30:22.31 & E-wf3 & 15.916  (0.010) \nl
293 & 626.353 & 706.461 & 5:25:44.472 & --69:30:19.09 & E-wf3 V & 13.132  (0.002) \nl
294 & 670.698 & 539.728 & 5:25:41.672 & --69:30:10.40 & E-wf3 V & 17.446  (0.038) \nl
295 & 682.269 & 146.000 & 5:25:34.596 & --69:29:58.77 & E-wf3 & 18.327  (0.041) \nl
296 & 750.581 & 194.599 & 5:25:35.816 & --69:29:53.59 & E-wf3 & 17.674  (0.028) \nl
297 & 206.881 & 139.141 & 5:25:32.027 & --69:30:44.03 & E-wf3 & 19.052  (0.058) \nl
298 & 236.359 & 150.349 & 5:25:32.378 & --69:30:41.46 & E-wf3 & 19.903  (0.118) \nl

300 & 117.117 & 139.786 & 5:25:30.292 & --69:31:08.99 & E-wf4 & 18.500  (0.037) \nl
301 & 102.388 & 661.826 & 5:25:27.260 & --69:31:58.27 & E-wf4 & 18.468  (0.040) \nl
302 & 151.012 & 732.273 & 5:25:27.777 & --69:32:06.37 & E-wf4 & 17.760  (0.028) \nl
303 & 141.534 & 319.154 & 5:25:29.764 & --69:31:26.75 & E-wf4 & 14.974  (0.004) \nl
304 & 207.554 & 416.185 & 5:25:30.445 & --69:31:37.84 & E-wf4 & 16.768  (0.011) \nl
305 & 202.375 & 632.783 & 5:25:29.216 & --69:31:58.32 & E-wf4 & 17.329  (0.017) \nl
306 & 234.848 & 587.563 & 5:25:30.031 & --69:31:54.93 & E-wf4 & 19.327  (0.076) \nl
307 & 241.896 & 607.643 & 5:25:30.072 & --69:31:57.06 & E-wf4 & 17.275  (0.016) \nl
308 & 221.568 & 338.525 & 5:25:31.113 & --69:31:30.78 & E-wf4 & 14.791  (0.004) \nl
309 & 251.783 & 170.844 & 5:25:32.572 & --69:31:15.62 & E-wf4 & 19.170  (0.064) \nl
310 & 302.518 & 549.339 & 5:25:31.467 & --69:31:53.20 & E-wf4 & 18.555  (0.041) \nl 
311 & 322.960 & 533.000 & 5:25:31.934 & --69:31:52.26 & E-wf4 & 18.695  (0.043) \nl
312 & 322.468 & 186.645 & 5:25:33.759 & --69:31:19.08 & E-wf4 & 15.599  (0.006) \nl
313 & 323.569 & 163.505 & 5:25:33.368 & --69:31:26.45 & E-wf4 & 17.680  (0.020) \nl
314 & 382.990 & 243.649 & 5:25:34.566 & --69:31:26.21 & E-wf4 & 19.120  (0.059) \nl
315 & 472.313 & 492.403 & 5:25:34.864 & --69:31:52.49 & E-wf4 & 19.702  (0.115) \nl
316 & 535.841 & 242.504 & 5:25:37.354 & --69:31:30.38 & E-wf4 & 19.307  (0.073) \nl
317 & 536.663 & 461.382 & 5:25:36.194 & --69:31:51.30 & E-wf4 & 18.316  (0.040) \nl
318 & 560.390 & 277.688 & 5:25:37.605 & --69:31:34.39 & E-wf4 & 18.847  (0.052) \nl
319 & 574.503 & 574.484 & 5:25:36.283 & --69:32:03.13 & E-wf4 & 19.150  (0.116) \nl
320 & 587.925 & 473.000 & 5:25:37.057 & --69:31:53.86 & E-wf4 & 18.696  (0.061) \nl
321 & 596.447 & 264.982 & 5:25:38.327 & --69:31:34.14 & E-wf4 & 18.586  (0.043) \nl
322 & 661.675 & 288.691 & 5:25:39.395 & --69:31:38.29 & E-wf4 & 13.836  (0.002) \nl
323 & 724.334 & 265.986 & 5:25:40.628 & --69:31:37.92 & E-wf4 & 18.924  (0.083) \nl
324 & 731.899 & 529.597 & 5:25:39.381 & --69:32:03.18 & E-wf4 V & 18.139  (0.070) \nl
325 & 749.995 & 672.309 & 5:25:38.940 & --69:32:17.20 & E-wf4 V & 14.710  (0.012) \nl
326 & 783.278 & 591.757 & 5:25:39.966 & --69:32:10.44 & E-wf4 V & 16.052  (0.026) \nl

330 & 694.271 & 083.784 & 5:25:26.769 & --69:31:25.29 & F-pc  & 15.792  (0.006) \nl
331 & 303.148 & 733.383 & 5:25:22.607 & --69:30:58.82 & F-pc  & 16.861  (0.011) \nl

340 & 306.655 & 070.736 & 5:25:24.597 & --69:30:45.41 & F-wf2 & 15.515  (0.005) \nl
341 & 473.100 & 057.589 & 5:25:21.543 & --69:30:41.28 & F-wf2 & 17.273  (0.013) \nl
342 & 531.532 & 611.436 & 5:25:23.891 & --69:29:47.36 & F-wf2 & 17.103  (0.014) \nl

350 & 130.376 & 721.599 & 5:25:41.565 & --69:31:10.63 & F-wf3 & 18.135  (0.028) \nl
351 & 152.138 & 726.227 & 5:25:41.783 & --69:31:08.75 & F-wf3 & 17.530  (0.018) \nl
352 & 156.710 & 613.613 & 5:25:39.810 & --69:31:04.74 & F-wf3 & 17.203  (0.013) \nl
353 & 231.375 & 385.683 & 5:25:36.202 & --69:30:50.37 & F-wf3 & 16.023  (0.007) \nl
354 & 262.401 & 634.561 & 5:25:40.843 & --69:30:55.54 & F-wf3 & 15.121  (0.005) \nl
355 & 612.542 & 613.529 & 5:25:42.638 & --69:30:22.01 & F-wf3 & 16.714  (0.013) \nl
356 & 675.392 & 695.695 & 5:25:44.477 & --69:30:18.82 & F-wf3 V & 13.843  (0.003) \nl
357 & 766.953 & 176.323 & 5:25:35.808 & --69:29:53.41 & F-wf3 & 16.662  (0.010) \nl
358 & 222.567 & 754.311 & 5:25:42.721 & --69:31:03.09 & F-wf3 & 16.104  (0.008) \nl
359 & 520.155 & 636.918 & 5:25:42.486 & --69:30:31.43 & F-wf3 & 17.808  (0.026) \nl
360 & 387.792 & 706.562 & 5:25:42.904 & --69:30:46.11 & F-wf3 & 17.656  (0.022) \nl
361 & 397.182 & 712.011 & 5:25:43.058 & --69:30:45.40 & F-wf3 & 17.932  (0.027) \nl
362 & 340.460 & 436.454 & 5:25:37.790 & --69:30:41.78 & F-wf3 & 17.162  (0.012) \nl

370 & 190.129 & 372.325 & 5:25:29.419 & --69:31:33.22 & F-wf4 & 13.918  (0.003) \nl
371 & 185.623 & 302.985 & 5:25:29.785 & --69:31:26.58 & F-wf4 & 15.736  (0.006) \nl
372 & 257.507 & 395.704 & 5:25:30.469 & --69:31:37.67 & F-wf4 & 17.499  (0.015) \nl
373 & 266.626 & 317.549 & 5:25:31.134 & --69:31:30.65 & F-wf4 & 15.502  (0.005) \nl
374 & 357.737 & 159.598 & 5:25:33.772 & --69:31:18.95 & F-wf4 & 16.368  (0.008) \nl
375 & 359.546 & 139.348 & 5:25:33.934 & --69:31:17.12 & F-wf4 & 16.820  (0.010) \nl
376 & 703.313 & 240.469 & 5:25:39.399 & --69:31:38.14 & F-wf4 & 14.585  (0.003) \nl
377 & 203.469 & 731.000 & 5:25:27.377 & --69:32:07.17 & F-wf4 & 16.609  (0.010) \nl

\enddata

\tablecomments{This table contains the raw measurements for all blinked and/or
checked objects. The objects are listed in the order they were measured and there are
many repeated observations for objects appearing on different images.}

\end{deluxetable}
 \newpage

\begin{deluxetable}{ccccccc}
\tablecaption{Far-UV photometry of field stars in the LMC }
\small
\tablehead{\colhead{Source no.} & \colhead{N$_{obs}$} &
 \colhead{ $\alpha$(J2000)}  
& \colhead{ $\delta$(J2000)} & \colhead{$<$FUV-mag$>$ (error) }}

\startdata
 26 &   1 &  5:25: 2.59 & --69:30:12.39 & 16.83  (0.01) \nl
 25 &   1 &  5:25: 4.62 & --69:30:20.24 & 15.24  (0.01) \nl
 23 &   2 &  5:25: 5.17 & --69:29:30.81 & 19.14  (0.16) \nl
  
 22 &   2 &  5:25: 5.61 & --69:29:33.39 & 18.73  (1.89) \nl
  
 21 &   2 &  5:25: 6.09 & --69:29:27.74 & 17.23  (0.05) \nl
  
 20 &   2 &  5:25: 7.32 & --69:30: 4.32 & 17.44  (0.13) \nl
  
  3 &   1 &  5:25: 8.10 & --69:30:35.52 & 18.69  (0.03) \nl
  5 &   2 &  5:25: 8.75 & --69:30:43.54 & 18.47  (0.08) \nl
   
117 &   1 &  5:25: 9.79 & --69:29:17.60 & 18.71  (0.08) \nl
  4 &   2 &  5:25: 9.89 & --69:30:41.32 & 13.20  (0.16) \nl
   
 19 &   2 &  5:25:10.77 & --69:30: 8.47 & 18.34  (0.16) \nl
  
116 &   1 &  5:25:11.42 & --69:29:31.95 & 18.01  (0.03) \nl
 54 &   1 &  5:25:11.61 & --69:31:39.53 & 18.95  (0.10) \nl
 15 &   2 &  5:25:11.73 & --69:30:10.14 & 18.40  (0.41) \nl

 55 &   1 &  5:25:12.07 & --69:31:36.14 & 18.29  (0.07) \nl
  8 &   1 &  5:25:12.19 & --69:31: 5.32 & 19.55  (0.08) \nl
  2 &   2 &  5:25:12.46 & --69:30:39.65 & 17.70  (0.04) \nl
   
  6 &   2 &  5:25:12.53 & --69:30:53.72 & 17.04  (0.16) \nl
   
 51 &   1 &  5:25:12.59 & --69:31:10.06 & 18.34  (0.06) \nl
  1 &   2 &  5:25:12.88 & --69:30:37.22 & 18.62  (0.18) \nl
   
 12 &   1 &  5:25:13.05 & --69:30:23.56 & 18.44  (0.03) \nl
 17 &   2 &  5:25:13.43 & --69:29:32.66 & 16.65  (0.10) \nl
  
 56 &   2 &  5:25:13.52 & --69:31:22.15 & 16.40  (0.01) \nl
  
 58 &   1 &  5:25:13.64 & --69:31:41.34 & 19.05  (0.43) \nl
  
 14 &   2 &  5:25:14.04 & --69:29:44.47 & 14.89  (0.11) \nl
  
 16 &   2 &  5:25:14.08 & --69:29:26.70 & 18.46  (0.14) \nl
103 &   1 &  5:25:14.09 & --69:30:34.72 & 18.23  (0.04) \nl
102 &   1 &  5:25:14.12 & --69:30:34.21 & 18.30  (0.04) \nl
115 &   1 &  5:25:14.33 & --69:29:28.46 & 18.28  (0.04) \nl
  7 &   1 &  5:25:14.51 & --69:30:50.67 & 16.19  (0.01) \nl
 18 &   1 &  5:25:14.73 & --69:29:10.86 & 16.79  (0.02) \nl
 59 &   2 &  5:25:14.77 & --69:31:24.61 & 18.49  (0.14) \nl
  
 13 &   2 &  5:25:14.94 & --69:29:49.07 & 15.07  (0.13) \nl
  
 60 &   2 &  5:25:15.29 & --69:31:18.25 & 17.98  (0.13) \nl
   
 53 &   3 &  5:25:15.38 & --69:30:43.18 & 13.75  (0.11) \nl
 
 52 &   2 &  5:25:15.42 & --69:30:39.98 & 18.71  (0.33) \nl
  
 64 &   2 &  5:25:15.67 & --69:31:51.87 & 15.86  (0.28) \nl
  
 10 &   2 &  5:25:15.81 & --69:30: 7.98 & 17.62  (0.08) \nl
  
 63 &   2 &  5:25:15.82 & --69:31:24.24 & 18.68  (0.34) \nl
  
 12 &   1 &  5:25:15.90 & --69:29:40.17 & 18.67  (0.04) \nl
 57 &   2 &  5:25:16.01 & --69:30:48.25 & 16.83  (0.12) \nl
   
101 &   1 &  5:25:16.10 & --69:30:17.35 & 18.71  (0.05) \nl
 11 &   2 &  5:25:16.43 & --69:29:41.14 & 15.81  (0.13) \nl
  
 67 &   2 &  5:25:16.59 & --69:31:48.99 & 18.29  (0.09) \nl
  
100 &   1 &  5:25:16.77 & --69:30:16.13 & 17.99  (0.03) \nl
114 &   1 &  5:25:16.97 & --69:29: 7.22 & 18.46  (0.05) \nl
 24 &   2 &  5:25:17.12 & --69:29:49.75 & 18.86  (0.14) \nl
  
 65 &   2 &  5:25:17.24 & --69:31:35.77 & 18.29  (0.13) \nl
  
 61 &   2 &  5:25:17.28 & --69:30:54.08 & 14.39  (0.07) \nl 
  
112 &   1 &  5:25:17.54 & --69:29:36.95 & 18.45  (0.04) \nl
111 &   1 &  5:25:17.62 & --69:29:34.38 & 17.90  (0.03) \nl 
 62 &   2 &  5:25:18.03 & --69:30:50.90 & 17.17  (0.12) \nl

113 &   1 &  5:25:18.43 & --69:29: 1.34 & 17.51  (0.03) \nl
110 &   1 &  5:25:19.21 & --69:29: 4.33 & 17.29  (0.02) \nl
 
 66 &   3 &  5:25:19.42 & --69:30:53.93 & 18.65  (0.16) \nl
  
 69 &   2 &  5:25:19.53 & --69:31:33.58 & 18.69  (0.10) \nl
  
 70 &   2 &  5:25:20.07 & --69:31:33.18 & 19.39  (0.29) \nl
   
 72 &   2 &  5:25:20.44 & --69:31:49.85 & 18.02  (0.19) \nl
  
 83 &   2 &  5:25:21.16 & --69:31:54.63 & 18.65  (0.08) \nl
  
 39 &   3 &  5:25:21.51 & --69:29:50.51 & 18.19  (0.69) \nl
  
 31 &   5 &  5:25:21.55 & --69:30:41.39 & 16.62  (0.42) \nl
  
259 &   1 &  5:25:21.71 & --69:29:37.33 & 15.64  (0.04) \nl
 
 71 &   3 &  5:25:21.78 & --69:31: 3.65 & 18.01  (0.15) \nl 
 68 &   2 &  5:25:21.79 & --69:30:56.33 & 19.80  (0.27) \nl
  
 74 &   1 &  5:25:21.86 & --69:31:54.43 & 22.53  (5.33) \nl
260 &   1 &  5:25:21.89 & --69:29:31.79 & 16.35  (0.19) \nl
 
 36 &   5 &  5:25:22.08 & --69:30:10.12 & 17.29  (0.61) \nl
   
 78 &   2 &  5:25:22.14 & --69:31:48.82 & 16.41  (0.06) \nl
  
 75 &   1 &  5:25:22.39 & --69:31:55.20 & 15.84  (0.11) \nl
  
 73 &   2 &  5:25:22.44 & --69:31:39.18 & 15.77  (0.10) \nl
  
331 &   1 &  5:25:22.61 & --69:30:58.82 & 16.48  (0.01) \nl
 82 &   2 &  5:25:22.71 & --69:31:44.86 & 18.14  (0.36) \nl
 
 40 &   3 &  5:25:22.79 & --69:29:51.16 & 19.15  (0.15) \nl
  
 33 &   4 &  5:25:23.22 & --69:30:40.48 & 17.22  (0.53) \nl
  
 76 &   2 &  5:25:23.33 & --69:31:54.73 & 15.20  (0.02) \nl
  
253 &   1 &  5:25:23.35 & --69:30:20.76 & 18.55  (0.05) \nl
130 &   1 &  5:25:23.42 & --69:29:28.92 & 19.28  (0.09) \nl
 
 44 &   6 &  5:25:23.89 & --69:29:47.48 & 16.30  (0.38) \nl 
  
 35 &   5 &  5:25:23.99 & --69:30:19.67 & 16.99  (0.53) \nl
  
 46 &   1 &  5:25:24.43 & --69:29:28.83 & 18.18  (0.04) \nl
142 &   1 &  5:25:24.48 & --69:31:38.04 & 18.17  (0.08) \nl
 32 &   4 &  5:25:24.59 & --69:30:45.32 & 14.85  (0.49) \nl
  
 43 &   2 &  5:25:24.65 & --69:29:47.48 & 17.50  (0.20) \nl
  
 77 &   2 &  5:25:25.37 & --69:31:43.37 & 14.94  (0.00) \nl
  
251 &   1 &  5:25:25.46 & --69:30: 6.04 & 19.06  (0.09) \nl
 79 &   1 &  5:25:25.60 & --69:31:46.04 & 15.92  (0.01) \nl
250 &   1 &  5:25:25.71 & --69:30: 6.51 & 19.66  (0.15) \nl
 42 &   2 &  5:25:25.92 & --69:29:50.48 & 17.42  (0.57) \nl
 
248 &   1 &  5:25:26.34 & --69:30:34.98 & 18.65  (0.06) \nl
 
 80 &   4 &  5:25:26.79 & --69:31:25.25 & 14.96  (0.34) \nl
38b &   2 &  5:25:26.86 & --69:30: 7.05 & 19.89  (0.52) \nl
 
38a &   3 &  5:25:26.95 & --69:30: 7.75 & 18.74  (0.33) \nl
 
37b &   2 &  5:25:27.08 & --69:30: 9.24 & 19.84  (1.02) \nl
 
168 &   1 &  5:25:27.13 & --69:30: 7.56 & 18.37  (0.05) \nl
 34 &   4 &  5:25:27.14 & --69:30:28.18 & 17.73  (0.46) \nl 
 
301 &   1 &  5:25:27.26 & --69:31:58.27 & 17.93  (0.04) \nl
377 &   1 &  5:25:27.38 & --69:32: 7.17 & 16.09  (0.01) \nl
 
37a &   5 &  5:25:27.43 & --69:30:10.07 & 18.05  (0.55) \nl
243 &   1 &  5:25:27.43 & --69:30:26.86 & 16.93  (0.02) \nl
 
302 &   1 &  5:25:27.78 & --69:32: 6.37 & 17.23  (0.03) \nl
 81 &   1 &  5:25:27.92 & --69:31: 9.27 & 18.95  (0.08) \nl
 45 &   1 &  5:25:27.99 & --69:29:44.84 & 20.68  (0.71) \nl

 41 &   2 &  5:25:28.27 & --69:29:57.50 & 18.63  (0.69) \nl
  
170 &   2 &  5:25:28.79 & --69:29:44.57 & 17.64  (0.06) \nl
 
135 &   1 &  5:25:28.83 & --69:29:10.50 & 16.75  (0.03) \nl
132 &   2 &  5:25:29.01 & --69:29:24.40 & 17.18  (0.46) \nl
 
241 &   1 &  5:25:29.03 & --69:30:26.59 & 18.42  (0.05) \nl
 
145 &   2 &  5:25:29.15 & --69:30:58.32 & 18.07  (0.26) \nl
305 &   1 &  5:25:29.22 & --69:31:58.32 & 16.79  (0.02) \nl
144 &   2 &  5:25:29.25 & --69:30:52.58 & 18.15  (0.42) \nl
 
370 &   1 &  5:25:29.42 & --69:31:33.22 & 13.40  (0.00) \nl
 
146 &   2 &  5:25:29.46 & --69:30:58.70 & 16.70  (0.35) \nl 
  
148 &   3 &  5:25:29.84 & --69:31:26.88 & 14.74  (0.38) \nl 
125 &   1 &  5:25:29.91 & --69:30: 1.78 & 18.73  (0.06) \nl
306 &   1 &  5:25:30.03 & --69:31:54.93 & 18.79  (0.08) \nl
307 &   1 &  5:25:30.07 & --69:31:57.06 & 16.74  (0.02) \nl
 
147 &   2 &  5:25:30.35 & --69:31: 9.18 & 18.20  (0.26) \nl
304 &   2 &  5:25:30.44 & --69:31:37.84 & 16.54  (0.37) \nl
 
129 &   2 &  5:25:31.09 & --69:29:52.35 & 17.84  (0.31) \nl
308 &   2 &  5:25:31.11 & --69:31:30.78 & 14.56  (0.36) \nl
 
310 &   1 &  5:25:31.47 & --69:31:53.20 & 18.02  (0.04) \nl
131 &   2 &  5:25:31.60 & --69:29:35.35 & 16.32  (0.41) \nl
 
133 &   2 &  5:25:31.78 & --69:29:22.96 & 15.98  (0.45) \nl
 
311 &   1 &  5:25:31.93 & --69:31:52.26 & 18.16  (0.04) \nl
297 &   1 &  5:25:32.03 & --69:30:44.03 & 18.52  (0.06) \nl
 
149 &   3 &  5:25:32.30 & --69:30:49.79 & 17.05  (0.31) \nl
298 &   1 &  5:25:32.38 & --69:30:41.46 & 19.37  (0.12) \nl
309 &   1 &  5:25:32.57 & --69:31:15.62 & 18.64  (0.06) \nl
134 &   1 &  5:25:32.75 & --69:29:17.87 & 18.57  (0.26) \nl
164 &   1 &  5:25:32.99 & --69:30:29.06 & 18.45  (0.05) \nl
128 &   1 &  5:25:33.09 & --69:29:58.83 & 17.94  (0.05) \nl
313 &   1 &  5:25:33.37 & --69:31:26.45 & 17.15  (0.02) \nl
 
213 &   3 &  5:25:33.72 & --69:31:19.20 & 15.25  (0.45) \nl
124 &   3 &  5:25:33.74 & --69:30:13.89 & 17.34  (0.39) \nl
 
162 &   1 &  5:25:33.81 & --69:30:31.64 & 18.87  (0.07) \nl
286 &   1 &  5:25:33.85 & --69:30:35.91 & 18.78  (0.07) \nl
214 &   2 &  5:25:33.88 & --69:31:17.06 & 16.63  (0.40) \nl
 
210 &   2 &  5:25:34.40 & --69:30:50.67 & 16.77  (0.03) \nl
123 &   1 &  5:25:34.51 & --69:30:16.64 & 17.36  (0.04) \nl
 
314 &   1 &  5:25:34.57 & --69:31:26.21 & 18.59  (0.06) \nl
295 &   1 &  5:25:34.60 & --69:29:58.77 & 17.79  (0.04) \nl
160 &   2 &  5:25:34.63 & --69:30:25.36 & 18.09  (0.07) \nl
211 &   1 &  5:25:34.66 & --69:30:48.19 & 17.91  (0.04) \nl
216 &   1 &  5:25:34.78 & --69:31:37.80 & 18.37  (0.06) \nl
315 &   1 &  5:25:34.86 & --69:31:52.49 & 19.17  (0.12) \nl
212 &   1 &  5:25:35.13 & --69:30:49.61 & 19.09  (0.09) \nl
296 &   2 &  5:25:35.82 & --69:29:53.59 & 16.53  (0.50) \nl
183 &   2 &  5:25:35.87 & --69:30:31.07 & 17.41  (0.06) \nl 
 
161 &   1 &  5:25:35.91 & --69:29:53.87 & 17.05  (0.02) \nl
 
215 &   3 &  5:25:36.19 & --69:30:50.29 & 14.89  (0.46) \nl
317 &   1 &  5:25:36.19 & --69:31:51.30 & 17.78  (0.04) \nl
 
180 &   2 &  5:25:36.22 & --69:30:38.05 & 18.41  (0.03) \nl
 
319 &   1 &  5:25:36.28 & --69:32: 3.13 & 18.61  (0.12) \nl
182 &   2 &  5:25:36.31 & --69:30:36.03 & 17.46  (0.07) \nl
218 &   1 &  5:25:36.31 & --69:31:39.65 & 18.43  (0.06) \nl 
 
186 &   1 &  5:25:36.48 & --69:30:17.29 & 19.21  (0.11) \nl
320 &   1 &  5:25:37.06 & --69:31:53.86 & 18.16  (0.06) \nl 
316 &   1 &  5:25:37.35 & --69:31:30.38 & 18.77  (0.07) \nl
184 &   2 &  5:25:37.41 & --69:30:29.90 & 17.43  (0.06) \nl
 
217 &   2 &  5:25:37.42 & --69:31: 5.48 & 17.44  (0.10) \nl
 
318 &   1 &  5:25:37.60 & --69:31:34.39 & 18.31  (0.05) \nl
283 &   2 &  5:25:37.77 & --69:30:42.00 & 16.16  (0.41) \nl
 
270 &   1 &  5:25:37.87 & --69:31: 8.27 & 17.26  (0.02) \nl
321 &   1 &  5:25:38.33 & --69:31:34.14 & 18.05  (0.04) \nl
202 &   1 &  5:25:38.46 & --69:30: 2.98 & 18.17  (0.04) \nl
282 &   1 &  5:25:38.70 & --69:30:41.51 & 18.33  (0.05) \nl
325 &   1 &  5:25:38.94 & --69:32:17.20 & 14.17  (0.01) \nl
221 &   3 &  5:25:39.36 & --69:31:38.42 & 13.48  (0.43) \nl
324 &   1 &  5:25:39.38 & --69:32: 3.18 & 17.60  (0.07) \nl
 
219 &   3 &  5:25:39.80 & --69:31: 4.67 & 16.03  (0.49) \nl
 
326 &   1 &  5:25:39.97 & --69:32:10.44 & 15.52  (0.03) \nl
205 &   1 &  5:25:40.10 & --69:29:50.34 & 17.37  (0.03) \nl
323 &   1 &  5:25:40.63 & --69:31:37.92 & 18.39  (0.08) \nl
207 &   1 &  5:25:40.77 & --69:29:41.24 & 18.25  (0.06) \nl
220 &   3 &  5:25:40.83 & --69:30:55.40 & 14.04  (0.42) \nl
 
272 &   1 &  5:25:41.08 & --69:31:11.26 & 18.38  (0.07) \nl
 
273 &   2 &  5:25:41.58 & --69:31:10.96 & 17.31  (0.27) \nl
204 &   1 &  5:25:41.61 & --69:29:52.55 & 16.84  (0.02) \nl
201 &   2 &  5:25:41.66 & --69:30:10.52 & 16.92  (0.01) \nl
 
222 &   3 &  5:25:41.77 & --69:31: 8.63 & 16.41  (0.45) \nl
 
187 &   1 &  5:25:42.44 & --69:30:23.55 & 17.33  (0.03) \nl
185 &   3 &  5:25:42.45 & --69:30:31.82 & 16.69  (0.46) \nl
 
188 &   3 &  5:25:42.60 & --69:30:22.35 & 15.63  (0.42) \nl 
203 &   1 &  5:25:42.61 & --69:29:57.64 & 13.35  (0.00) \nl
 
358 &   1 &  5:25:42.72 & --69:31: 3.09 & 15.59  (0.01) \nl
181 &   3 &  5:25:42.88 & --69:30:46.50 & 16.60  (0.42) \nl
 
280 &   1 &  5:25:42.97 & --69:30:55.47 & 17.40  (0.04) \nl  
223 &   1 &  5:25:42.70 & --69:31:17.83 & 18.26  (0.05) \nl
361 &   1 &  5:25:43.06 & --69:30:45.40 & 17.41  (0.03) \nl
224 &   1 &  5:25:43.09 & --69:31:17.91 & 18.56  (0.07) \nl
227 &   1 &  5:25:44.03 & --69:31:52.10 & 16.34  (0.05) \nl
225 &   1 &  5:25:44.32 & --69:31:19.30 & 17.93  (0.05) \nl
200 &   1 &  5:25:44.37 & --69:30:16.89 & 17.19  (0.03) \nl
189 &   3 &  5:25:44.45 & --69:30:19.17 & 12.89  (0.36) \nl
 
228 &   1 &  5:25:44.91 & --69:31:40.36 & 17.73  (0.09) \nl 
229 &   1 &  5:25:45.51 & --69:31:20.20 & 15.32  (0.01) \nl
206 &   1 &  5:25:45.90 & --69:29:57.45 & 15.72  (0.01) \nl
226 &   1 &  5:25:46.41 & --69:31: 4.39 & 14.78  (0.01) \nl
  
\enddata

\tablecomments{This table contains the average measurements for all objects 
in Table 2. Star 77 appears on two images, but only one measurement yielded a
UV magnitude. Star 43 has two measurements which are separated in position
by slighly more than 0.5 arcsec; we considered them to refer to the same object
because the UV magnitudes were very similar, despite the positional
difference.}

\end{deluxetable}

\newpage

\begin{deluxetable}{rrrr}
\tablecaption{LMC UV objects with UIT counterparts}
\small
\tablehead{\colhead{UIT ID} & \colhead{UIT B1} &
\colhead{HST counterpart} & \colhead{HST UV mag.}}
\startdata
15414 & 14.06 (0.04) & 25	& 15.24 (0.01) \nl
15511 & 11.47 (0.08) & 4	& 13.20 (0.16) \nl
15543 & 14.59 (0.06) & 6	& 17.04 (0.16) \nl
15556 & 15.05 (0.05) & 17	& 16.65 (0.10) \nl
15557 & 14.90 (0.03) & 56	& 16.40 (0.01) \nl
15565 & 13.35 (0.04) & 14 	& 14.89 (0.11) \nl
15584 & 13.58 (0.04) & 13 	& 15.07 (0.13) \nl
      &              & 24	& 18.86 (0.14) \nl
15591 & 12.20 (0.03) & 53	& 13.75 (0.11) \nl
15593 & 14.31 (0.04) & 64	& 15.86 (0.28) \nl
      &              & 67	& 18.29 (0.09) \nl
15598 & 14.25 (0.03) & 11	& 15.81 (0.13) \nl
      &		     & 12	& 18.67 (0.04) \nl
15602 & 12.79 (0.03) & 114 	& 18.46 (0.05) \nl
15616 & 12.83 (0.02) & 61	& 14.39 (0.07) \nl
      &		     & 62	& 17.17 (0.12) \nl
      &		     & 66	& 18.65 (0.16) \nl
15681 & 15.35 (0.04) & 31	& 16.62 (0.42) \nl
      &		     & 33	& 17.22 (0.53) \nl
15697 & 14.30 (0.04) & 73 	& 15.77 (0.10) \nl
15709 & 13.86 (0.05) & 76 	& 15.20 (0.02) \nl
      &	     	     & 74 	& 22.53 (5.33) \nl
      &	     	     & 75 	& 15.84 (0.11) \nl
15721 & 15.32 (0.06) & 44	& 16.30 (0.38) \nl
      &		     & 43	& 17.50 (0.20) \nl
15731 & 13.64 (0.04) & 32	& 14.85 (0.49) \nl
15744 & 13.06 (0.03) & 77	& 14.94 (0.00) \nl
      &		     & 79 	& 15.92 (0.01) \nl
15777 & 13.79 (0.04) & 80 	& 14.96 (0.34) \nl
15810 & 15.25 (0.05) & 135	& 16.75 (0.03) \nl
15821 & 15.54 (0.05) & 146	& 16.70 (0.35) \nl
      &		     & 145	& 18.07 (0.26) \nl
      &		     & 144	& 18.15 (0.42) \nl
15835 & 13.74 (0.04) & 148	& 14.74 (0.38) \nl
15853 & 15.71 (0.09) & 304	& 16.54 (0.37) \nl
15865 & 13.65 (0.04) & 308	& 14.56 (0.36) \nl
15873 & 15.11 (0.05) & 133 	& 15.98 (0.45) \nl
15906 & 16.53 (0.16) & 134 	& 18.57 (0.26) \nl
15922 & 14.25 (0.03) & 213 	& 15.25 (0.45) \nl
      &	     	     & 214	& 16.63 (0.40) \nl
15933 & 16.08 (0.12) & 210	& 16.77 (0.03) \nl
      &		     & 211	& 17.91 (0.04) \nl
      & 	     & 212	& 19.09 (0.09) \nl
15953 & 14.04 (0.06) & 215 	& 14.89 (0.46) \nl
15978 & 15.02 (0.04) & 283	& 16.16 (0.41) \nl
      &		     & 282	& 18.33 (0.05) \nl
15988 & 14.16 (0.04) & 325	& 14.17 (0.01) \nl
15995 & 12.82 (0.03) & 221	& 13.48 (0.43) \nl
      &		     & 323	& 18.39 (0.08) \nl
16004 & 14.99 (0.05) & 219	& 16.03 (0.49) \nl
16005 & 15.39 (0.06) & 326	& 15.52 (0.03) \nl
16019 & 13.23 (0.03) & 220	& 14.04 (0.42) \nl
      &		     & 280	& 17.40 (0.04) \nl
16038 & 14.98 (0.05) & 222	& 16.41 (0.45) \nl
      &		     & 272	& 18.38 (0.07) \nl
      &		     & 273	& 17.31 (0.27) \nl
16049 & 14.38 (0.05) & 188	& 15.63 (0.42) \nl
      &		     & 187	& 17.33 (0.03) \nl
16052 & 12.76 (0.03) & 203 	& 13.35 (0.00) \nl
16053 & 15.37 (0.05) & 181	& 16.60 (0.42) \nl
      &		     & 361	& 17.41 (0.03) \nl
16076 & 14.15 (0.41) & 189	& 12.89 (0.36) \nl
      &		     & 225	& 17.93 (0.05) \nl
16087 & 15.48 (0.40) & 200	& 17.19 (0.03) \nl
16096 & 14.74 (0.04) & 206	& 15.72 (0.01) \nl
\enddata
\tablecomments{The table contains information about those UIT objects
for which we found UV counterparts in the HST WFPC2 mosaic. We list
for each source more than one object, including those stars from Table
3 which are probably contained within the UIT photometry aperture
of the single UV source listed.}

\end{deluxetable}

\newpage

\begin{deluxetable}{crr}
\tablecaption{LMC UV objects with ROE/NRL counterparts}
\small
\tablehead{\colhead{Number} & \colhead{UV} &
\colhead{UV--V} }
\startdata
26	&	16.83$\pm$0.01	&	-2.59$\pm$0.04 \nl
25	&	15.24$\pm$0.01	&	-5.03$\pm$0.04 \nl
20	&	17.44$\pm$0.13	&	-1.85$\pm$0.16 \nl
4	&	13.20$\pm$0.16	&	-5.18$\pm$0.18 \nl
6	&	17.04$\pm$0.16	&	-2.32$\pm$0.18 \nl
12	&	18.44$\pm$0.03	&	-0.30$\pm$0.08 \nl
56	&	16.40$\pm$0.01	&	-4.60$\pm$0.04 \nl
53	&	13.75$\pm$0.11	&	-5.09$\pm$0.15 \nl
64	&	15.86$\pm$0.28	&	-4.24$\pm$0.24 \nl
11	&	15.81$\pm$0.13	&	-5.04$\pm$0.16 \nl
114	&	18.46$\pm$0.05	&	-1.10$\pm$0.10 \nl
61	&	14.39$\pm$0.07	&	-4.42$\pm$0.12 \nl
69	&	18.69$\pm$0.10	&	-0.65$\pm$0.14 \nl
68	&	19.80$\pm$0.27	&	 1.10$\pm$0.23 \nl
36	&	17.29$\pm$0.61	&	-2.83$\pm$0.35 \nl
78	&	16.41$\pm$0.06	&	-2.24$\pm$0.01 \nl
73	&	15.77$\pm$0.10	&	-3.80$\pm$0.14 \nl
331	&	16.48$\pm$0.01	&	-2.71$\pm$0.04 \nl
76	&	15.20$\pm$0.02	&	-5.08$\pm$0.06 \nl
46	&	18.18$\pm$0.04	&	-2.82$\pm$0.09 \nl
32	&	14.85$\pm$0.49	&	-6.00$\pm$0.31 \nl
77	&	14.94$\pm$0.00	&	-3.95$\pm$0.01 \nl
80	&	14.96$\pm$0.34	&	-4.22$\pm$0.26 \nl
3702	&	19.84$\pm$1.02	&	 0.90$\pm$0.45 \nl
168	&	18.37$\pm$0.05	&	 1.80$\pm$0.10 \nl
302	&	17.23$\pm$0.03	&	-2.09$\pm$0.08 \nl
81	&	18.95$\pm$0.08	&	 0.50$\pm$0.13 \nl
45	&	20.68$\pm$0.71	&	 1.83$\pm$0.38 \nl
45	&	18.33$\pm$0.07	&	-2.20$\pm$0.12 \nl
170	&	17.64$\pm$0.06	&	-2.85$\pm$0.11 \nl
305	&	16.79$\pm$0.02	&	-1.34$\pm$0.06 \nl
146	&	16.70$\pm$0.35	&	-4.30$\pm$0.26 \nl
148	&	14.74$\pm$0.38	&	-5.10$\pm$0.28 \nl
304	&	16.54$\pm$0.37	&	-3.28$\pm$0.27 \nl
308	&	14.56$\pm$0.36	&	-5.05$\pm$0.27 \nl
131	&	16.32$\pm$0.41	&	-4.65$\pm$0.29 \nl
133	&	15.98$\pm$0.45	&	-2.70$\pm$0.30 \nl
134	&	18.57$\pm$0.26	&	 0.44$\pm$0.23 \nl
313	&	17.15$\pm$0.02	&	-2.88$\pm$0.06 \nl
213	&	15.25$\pm$0.45	&	-3.80$\pm$0.30 \nl
212	&	19.09$\pm$0.09	&	-2.05$\pm$0.13 \nl
215	&	14.89$\pm$0.46	&	-5.00$\pm$0.30 \nl
319	&	18.61$\pm$0.12	&	-2.25$\pm$0.15 \nl
182	&	17.46$\pm$0.07	&	-1.63$\pm$0.12 \nl
186	&	19.21$\pm$0.11	&	-0.10$\pm$0.15 \nl
316	&	18.77$\pm$0.07	&	-0.01$\pm$0.12 \nl
221	&	13.48$\pm$0.43	&	-5.40$\pm$0.29 \nl
326	&	15.52$\pm$0.03	&	-3.95$\pm$0.08 \nl
205	&	17.37$\pm$0.03	&	-2.00$\pm$0.08 \nl
220	&	14.04$\pm$0.42	&	-4.65$\pm$0.29 \nl

273	&	17.31$\pm$0.27	&	-2.88$\pm$0.23 \nl
201	&	16.92$\pm$0.01	&	-3.40$\pm$0.04 \nl
188	&	15.63$\pm$0.42	&	-4.90$\pm$0.29 \nl
203	&	13.35$\pm$0.00	&	-5.00$\pm$0.01 \nl
223	&	18.26$\pm$0.05	&	-0.60$\pm$0.10 \nl
200	&	17.19$\pm$0.03	&	-0.90$\pm$0.08 \nl
189	&	12.89$\pm$0.36	&	-5.70$\pm$0.27 \nl
226	&	14.78$\pm$0.01	&	-4.34$\pm$0.04 \nl 
\enddata
\tablecomments{The V magnitude used in the UV--V color index is transformed
from the ``short-R'' magnitude given in the ROE/NRL catalog. }

\end{deluxetable}
 \newpage

\begin{figure}[tbh]
\vspace{14cm}
\includegraphics{LMCmos4_ov.ps}
\end{figure} 
   
\newpage

\begin{figure}[tbh]
\vspace{14cm}
\includegraphics{LMC_hist1A.ps}
\end{figure}
   
\newpage
 
\begin{figure}[htb]
\putplot{singleobs.eps}{1in}{-90}{60}{60}{-250}{55}
\putplot{multiobs.eps}{1in}{-90}{60}{60}{-250}{-102}
\end{figure}

\newpage
  
\begin{figure}[tbh]
\vspace{14cm}
\includegraphics{LMC_plot1A.ps}
\end{figure}

\newpage

\begin{figure}[tbh]
\vspace{14cm}
\includegraphics{LMC-hist2.eps}
\end{figure}

\newpage
   
\begin{figure}[tbh]
\vspace{14cm}
\includegraphics{LMC_Faust_cm.eps}
\end{figure} 

\newpage

\begin{figure}[tbh]
\vspace{14cm}
\includegraphics{LMC-UIT-lumFN.eps}
\end{figure}

\newpage

\begin{figure}[tbh]
\vspace{14cm}
\includegraphics{LMC-UIT-logLumFN.eps}
\end{figure}

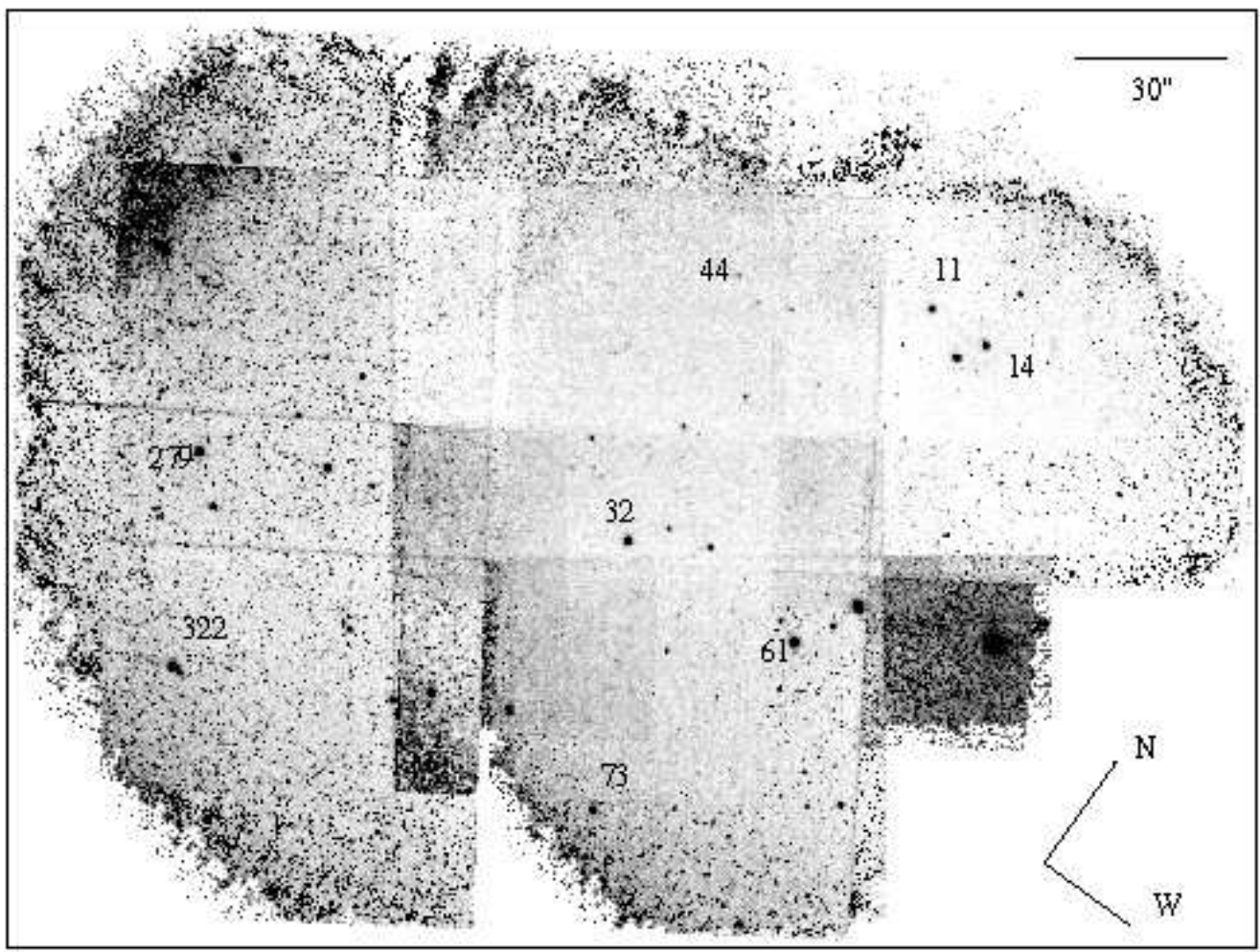
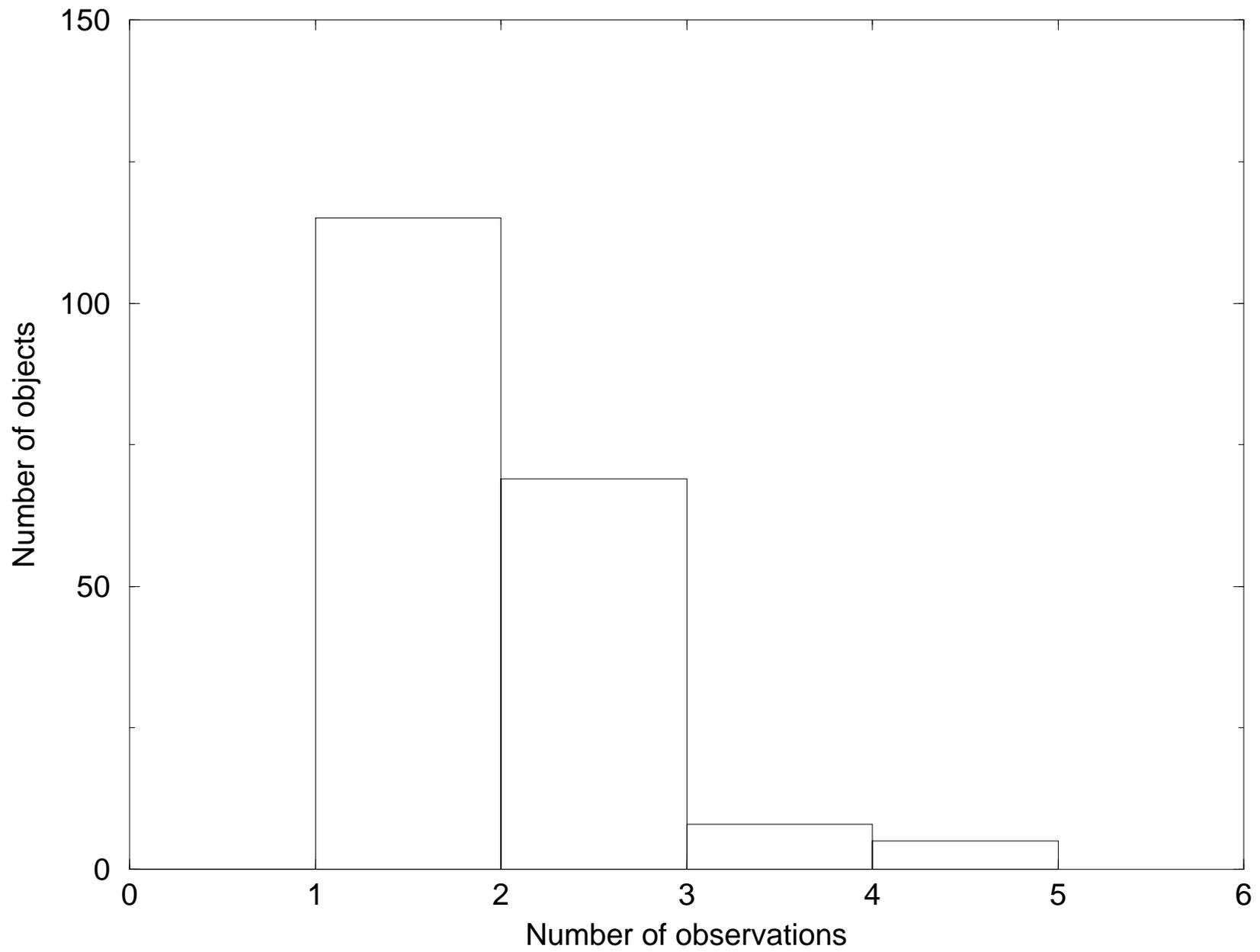
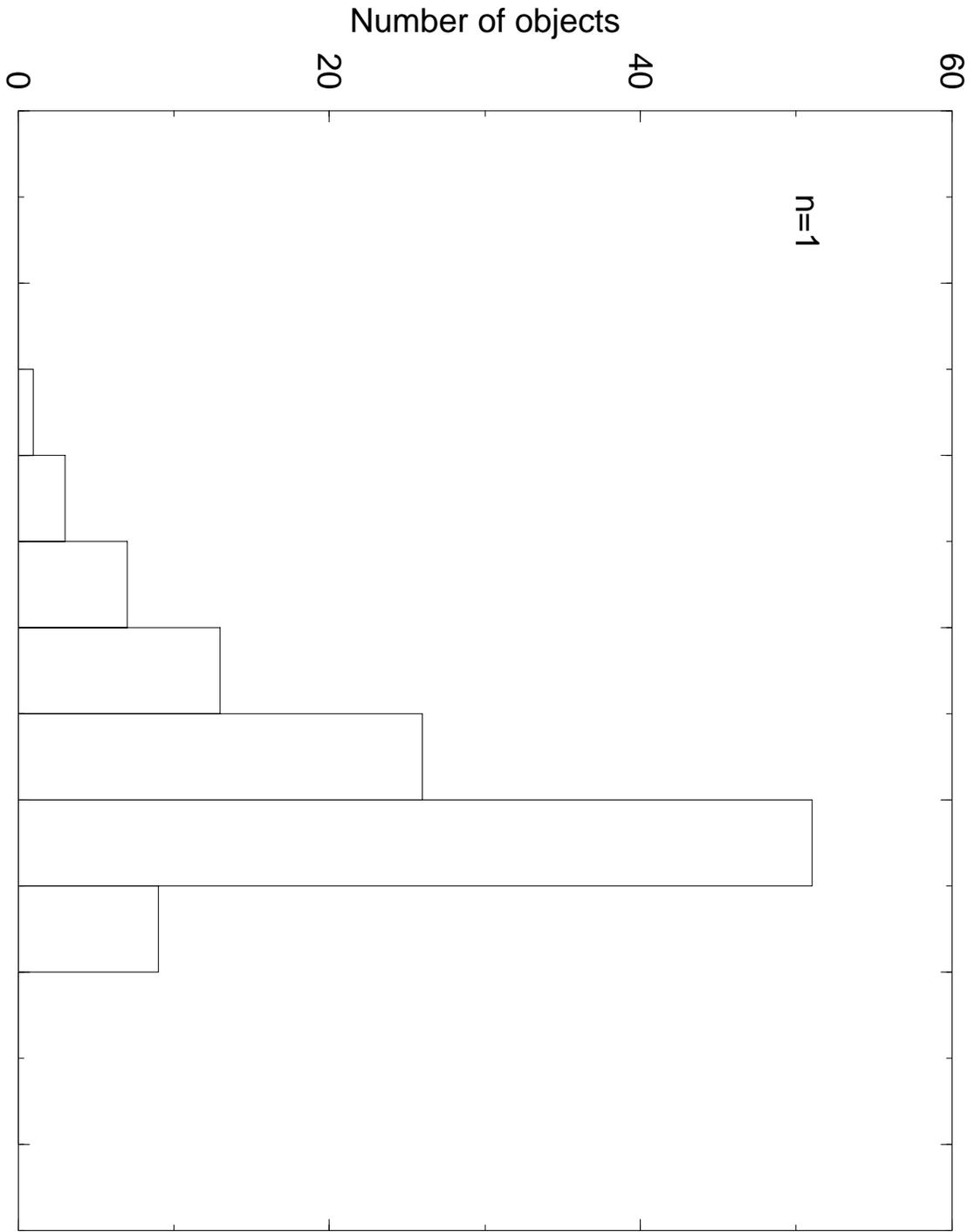
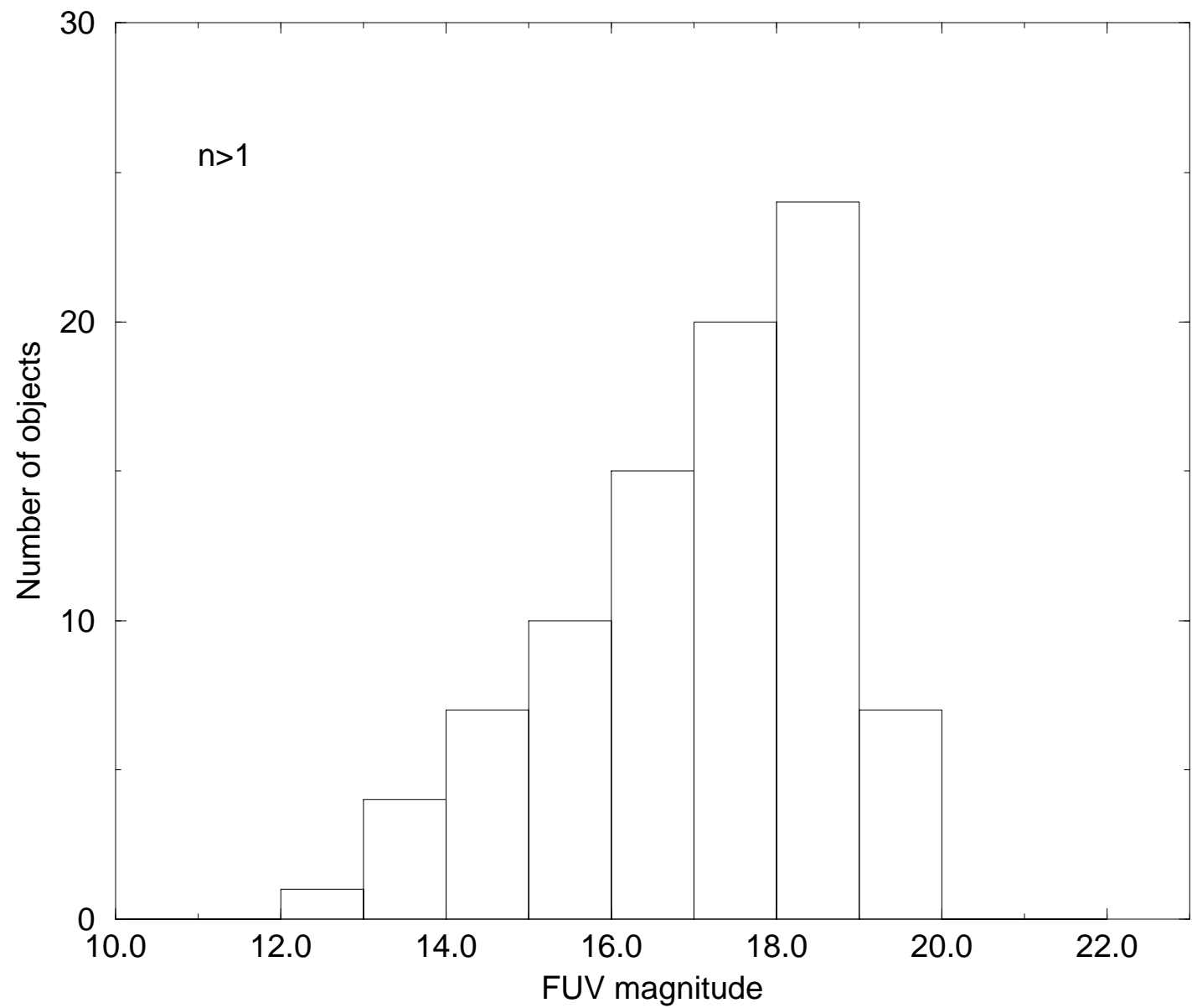
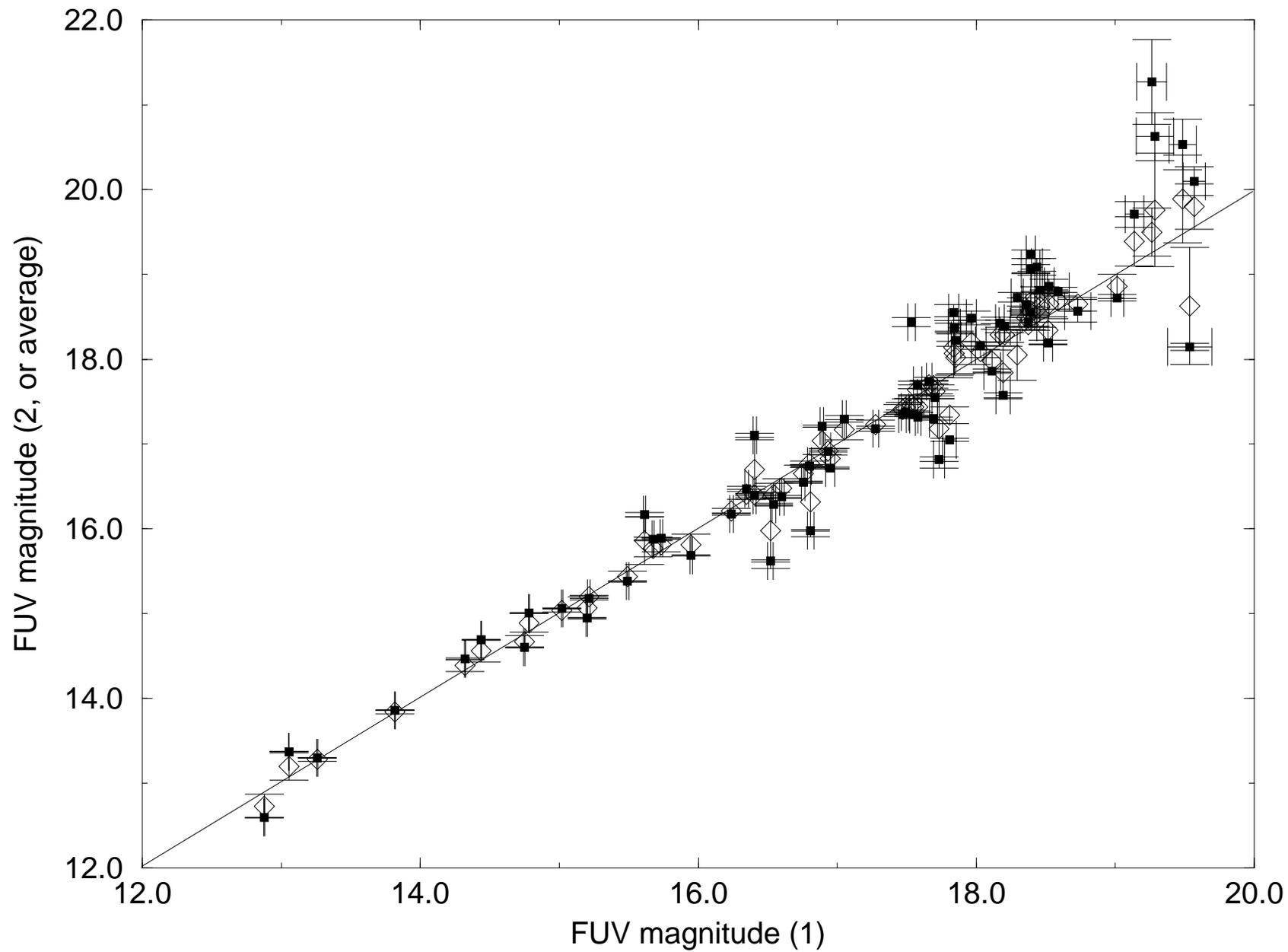
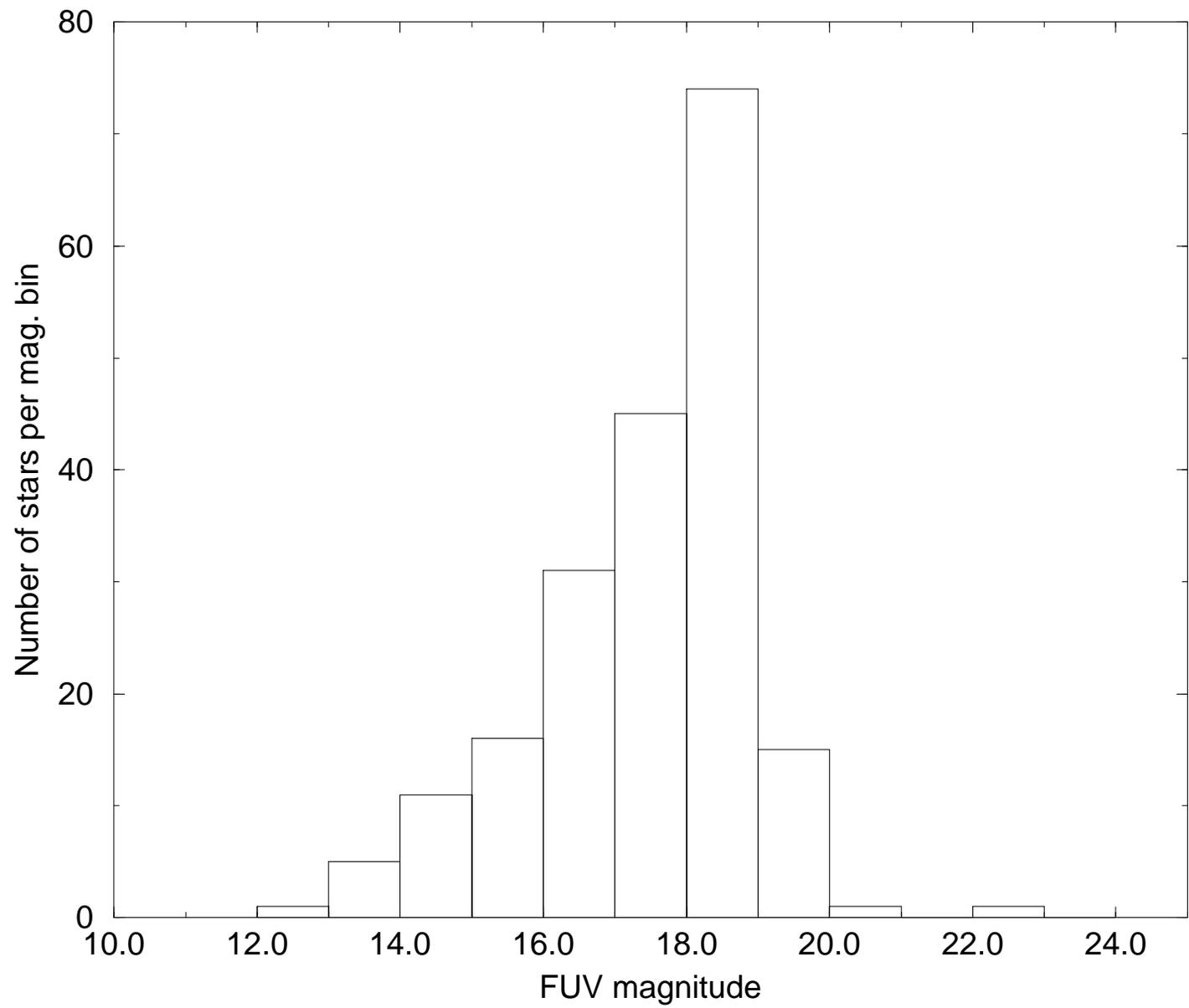
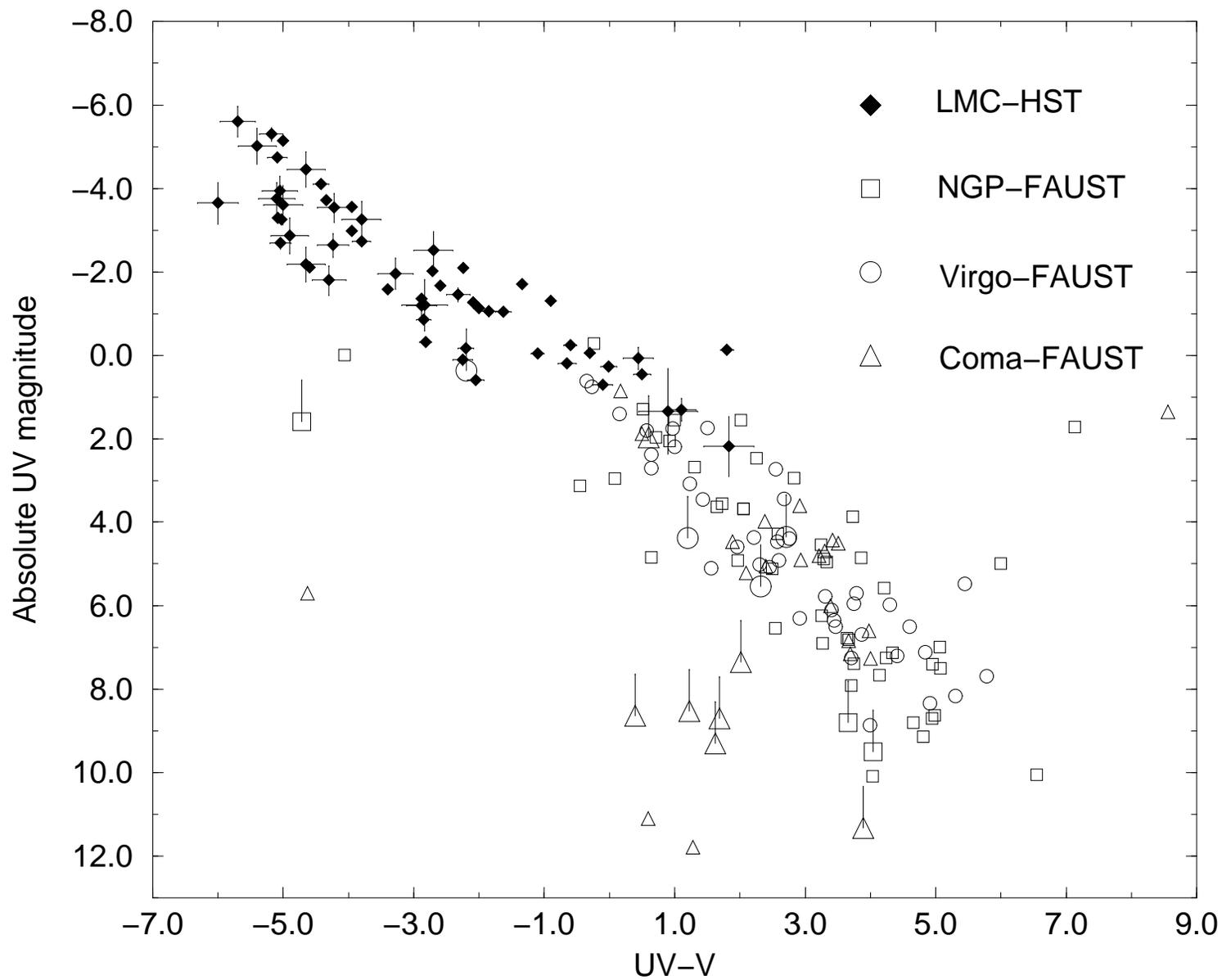
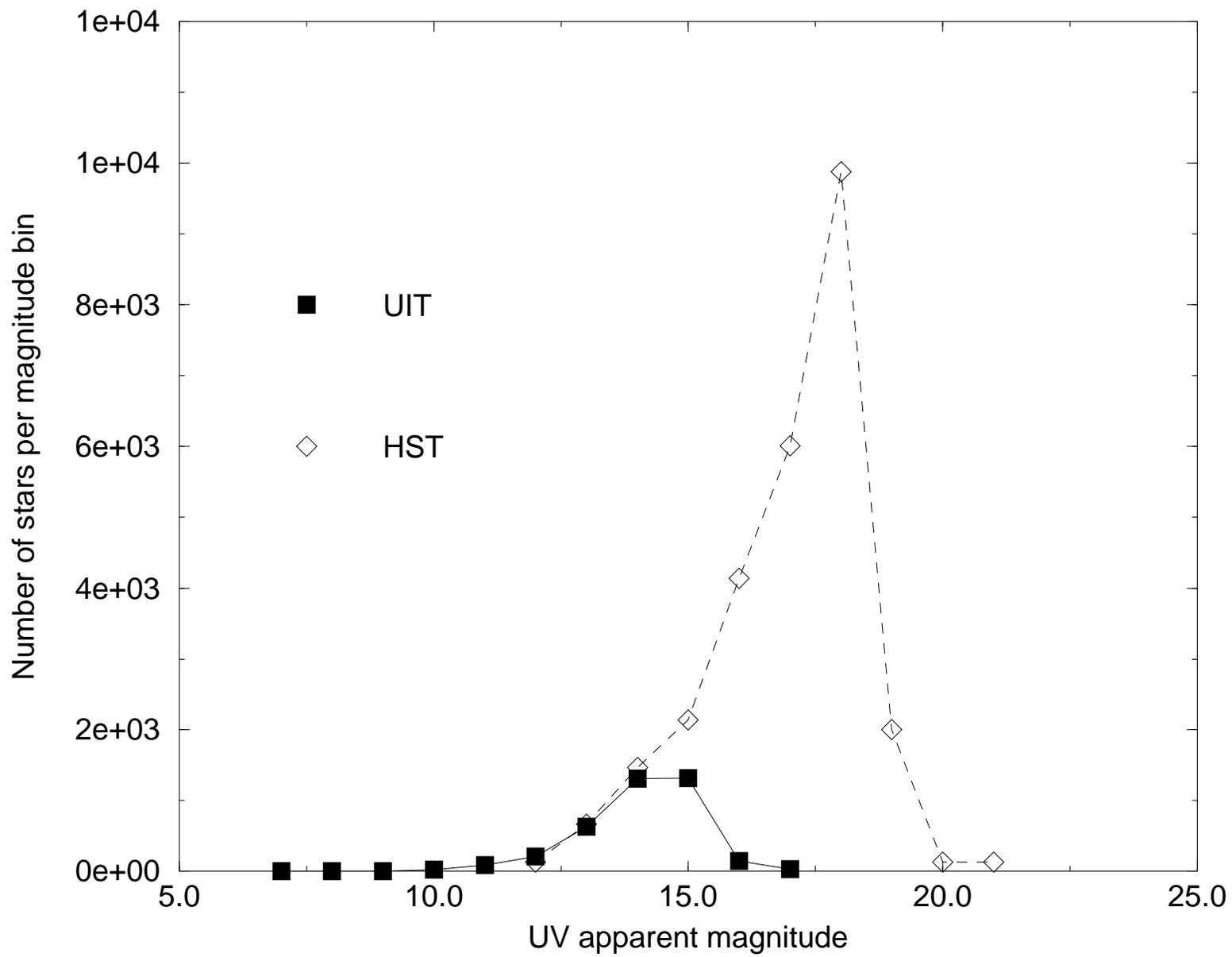
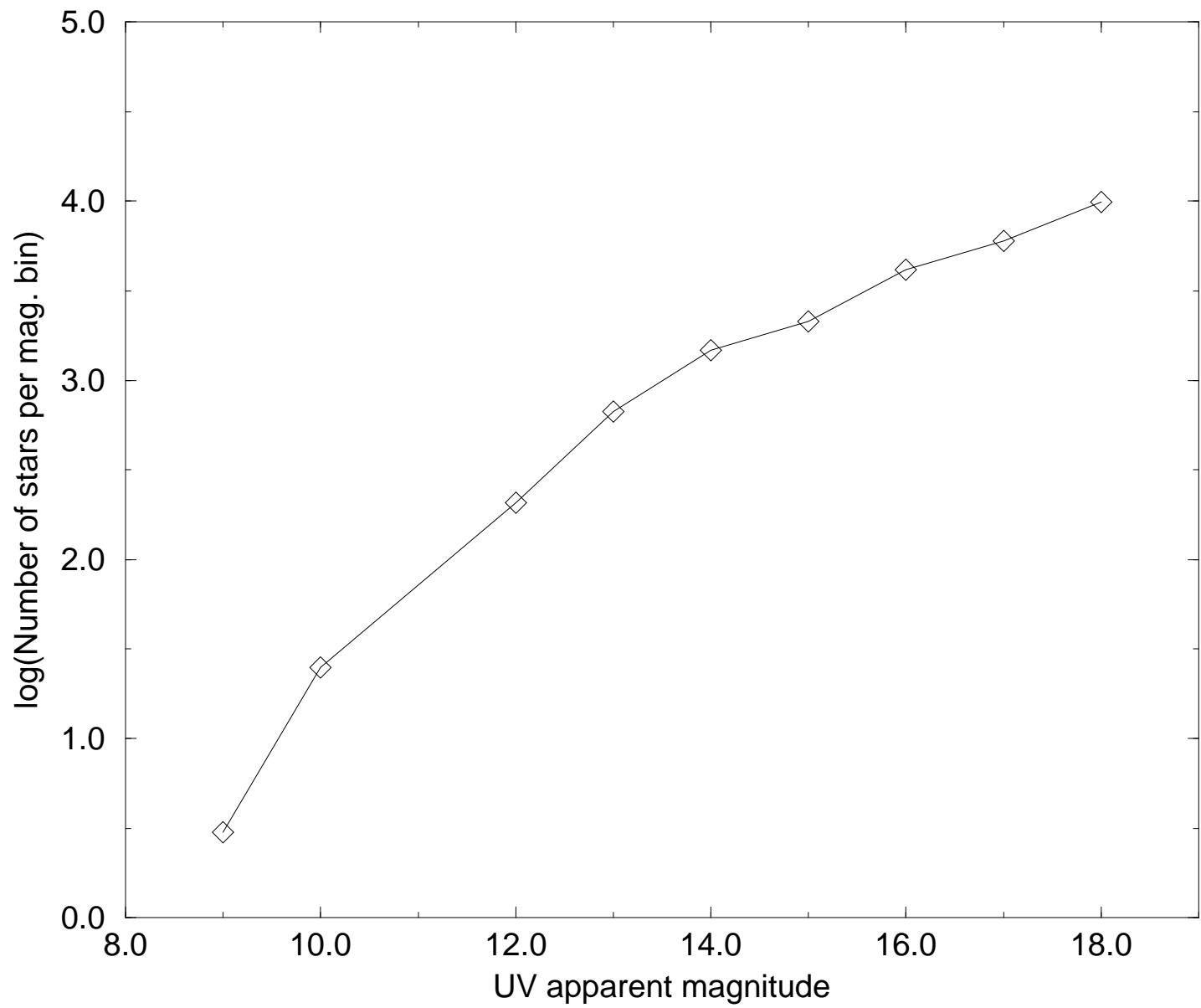
\end{document}